\documentclass[journal,letterpaper,twoside]{IEEEtran_nssmic}
\usepackage{graphics}
\usepackage{graphicx}
\usepackage{cite}
\usepackage{subfigure}
\usepackage{epsfig}
\begin{document}

% Title, authors and addresses

% use the thanksref command within \title, \author or \address for footnotes;
% use the corauthref command within \author for corresponding author footnotes;
% use the ead command for the email address,
% and the form \ead[url] for the home page:
% \title{Title\thanksref{label1}}
% \thanks[label1]{}
% \author{Name\corauthref{cor1}\thanksref{label2}}
% \ead{email address}
% \ead[url]{home page}
% \thanks[label2]{}
% \corauth[cor1]{}
% \address{Address\thanksref{label3}}
% \thanks[label3]{}

\title{Simulation of Heavily Irradiated Silicon Pixel Sensors and Comparison with Test Beam Measurements}
% use optional labels to link authors explicitly to addresses:
% \author[label1,label2]{}
% \address[label1]{}
% \address[label2]{}
%
% Author list
%
%\author[1]{Vincenzo~Chiochia,~\IEEEmembership{IEEE,~Member,}%
%\corauth[1]{Corresponding author. Address: CERN, CH-1211 Gen\'eve 23, Switzerland. Phone: 0041-22-7676041, Fax: 0041-22-7678340}}
\author{Vincenzo~Chiochia$^*$,~\IEEEmembership{IEEE,~Member,}
        Morris~Swartz,
        Daniela~Bortoletto,
	Lucien~Cremaldi,~\IEEEmembership{IEEE,~Member,}
        Susanna Cucciarelli,
        Andrei~Dorokhov,
	Christoph~H\"ormann,
	Dongwook~Kim,
        Marcin~Konecki,
	Danek~Kotlinski,
        Kirill~Prokofiev,
        Christian~Regenfus,
	Tilman~Rohe,~\IEEEmembership{IEEE,~Member,}
        David~A.~Sanders,~\IEEEmembership{IEEE,~Member,}
        Seunghee~Son,
        Thomas~Speer%
%\thanks{Manuscript received January 20, 2002; revised August 13, 2002.
%        This work was supported by the IEEE.}% <-this % stops a space
\thanks{$^*$Corresponding author: Address: CERN, CH-1211 Gen\'eve 23, Switzerland. Phone: 0041-22-7676041, Fax: 0041-22-7678340, e-mail: vincenzo.chiochia@cern.ch.}
\thanks{V.~Chiochia, C.~Regenfus and T.~Speer are with the Physik Institut der Universit\"at Z\"urich, 8057 Z\"urich, Switzerland.}
\thanks{M.~Swartz and D.~Kim are with the Johns Hopkins University, Baltimore, MD 21218, USA.}
\thanks{D.~Bortoletto and S.~Son are with the Purdue University, Task G, West Lafayette, IN 47907, USA.}
\thanks{L.~Cremaldi and D.A.~Sanders are with the University of Mississippi, Department of Physics and Astronomy, University, MS 38677, USA}
\thanks{S.~Cucciarelli and M.~Konecki are with the Institut f\"ur Physik der Universit\"at Basel, 4056 Basel, Switzerland.}
\thanks{C.~H\"ormann, A.~Dorokhov and K.~Prokofiev are with the Physik Institut der Universit\"at Z\"urich, 8057 Z\"urich, Switzerland and the Paul Scherrer Institut, 5232 Villigen, Switzerland.}
\thanks{D.~Kotlinski and T.~Rohe are with the Paul Scherrer Institut, 5232 Villigen, Switzerland.}
}
%
%\corauth[1]{Corresponding author. Address: CERN, CH-1211 Gen\'eve 23, Switzerland. Phone: 0041-22-7676041, Fax: 0041-22-7678340}
\pubid{0000--0000/00\$00.00~\copyright~2005 IEEE}

\maketitle
%----------------------------------------
% Abstract
%----------------------------------------
\begin{abstract}
Charge collection measurements performed on heavily irradiated p-spray DOFZ pixel sensors with a grazing angle hadron beam provide a sensitive determination of the electric field within the detectors.  The data are compared with a complete charge transport simulation of the sensor which includes free carrier trapping and charge induction effects.  A linearly varying electric field based upon the standard picture of a constant type-inverted effective doping density is inconsistent with the data.  A two-trap double junction model implemented in the ISE TCAD software can be tuned to produce a double peak electric field which describes the data reasonably well. The modeled field differs somewhat from previous determinations based upon the transient current technique.  The model can also account for the level of charge trapping observed in the data.
\end{abstract}
%
%-----------------------------------------
% Introduction
%-----------------------------------------
\section{Introduction}
\label{sec:intro}

\PARstart{I}{n} the recent years, detectors consisting of one and two dimensional arrays of silicon diodes have come into widespread use as tracking detectors in particle and nuclear physics experiments.  It is well understood that the intra-diode electric fields in these detectors vary linearly in depth reaching a maximum value at the p-n junction.  The linear behavior is a consequence of a constant space charge density, $N_{\rm eff}$, caused by thermodynamically ionized impurities in the bulk material.  It is well known that the detector characteristics are affected by radiation exposure, but it is generally assumed that the same picture is valid after irradiation.  In fact, it is common to characterize the effects of irradiation in terms of a varying effective charge density.  The use of $N_{\rm eff}$ to characterize radiation damage has persisted despite a growing body of evidence \cite{lk}-\cite{castaldini} that the electric field does not vary linearly as a function of depth after heavy irradiation but instead exhibits maxima at both n$^+$ and p$^+$ implants.  The study presented in this paper demonstrates conclusively that the standard picture does not provide a good description of irradiated silicon pixel detectors.  We show that it is possible to adequately describe the charge collection characteristics of a heavily irradiated silicon detector in terms of a tuned double junction model which produces a double peak electric field profile across the detector.  The allowed parameter space of the model can also accommodate the expected level of leakage current and the level of charge trapping observed in the detector.

This paper is organized as follows: Section~\ref{sec:technique} describes the experimental technique and data, Section~\ref{sec:sim} describes the carrier transport simulation used to interpret the data, Section~\ref{sec:djmodels} describes the technique used to model double peak electric fields and the limitations of previous models. The tuning of a successful model is discussed in Section~\ref{sec:results}. Section~\ref{sec:conclusions} summarizes the results and develops several conclusions.
\pubidadjcol

%-------------------------------------------------
% Sec 2. Experimental technique and data
%-------------------------------------------------
\section{Experimental Technique and Data}
\label{sec:technique}

\begin{figure}[hbt]
%
% Figure 1
%
  \begin{center}
    \resizebox{\linewidth}{!}{\includegraphics{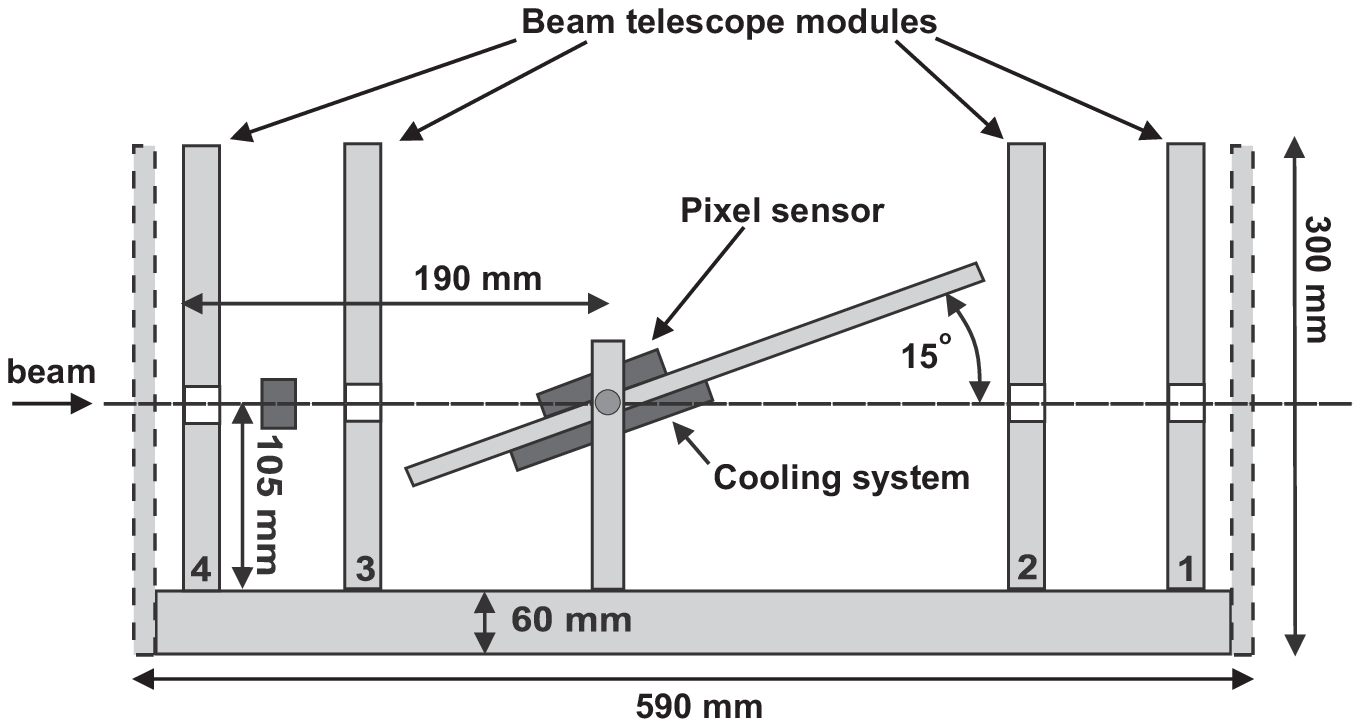}}
  \caption{Side view of the beam test apparatus consisting of four horizontal and four vertical planes of 
silicon strip detectors and a rotating stage for the pixel detector.}
  \label{fig:testbeam} 
  \end{center}
\end{figure}

This investigation is based upon beam test data that were accumulated as part of a program to develop a silicon pixel tracking detector \cite{tdr} for the CMS experiment at the CERN Large Hadron Collider.  The measurements were performed in the H2 line of the CERN SPS in 2003/04 using 150-225 GeV pions.  The beam test apparatus is described in~\cite{andrei} and is shown in Fig.~\ref{fig:testbeam}.  
A silicon beam telescope \cite{telescope} consisted of four modules each containing two 300~$\mu$m thick single-sided silicon detectors with a strip pitch of 25 $\mu$m and readout pitch of 50~$\mu$m.  The two detectors in each module were oriented to measure horizontal and vertical impact coordinates.  A pixel hybrid detector was mounted between the second and third telescope modules on a cooled rotating stage.  A trigger signal was generated by a silicon PIN diode. The analog signals from all detectors were digitized in a VME-based readout system by two CAEN (V550) and one custom built flash ADCs. The entire assembly was located in an open-geometry 3T Helmholtz magnet that produced a magnetic field parallel or orthogonal to the beam. The temperature of the test sensors was controlled with a Peltier cooler that was capable of operating down to -30$^\circ$C.  The telescope information was used to reconstruct the trajectories of individual beam particles and to achieve a precise determination of the particle hit position in the pixel detector.  The resulting intrinsic resolution of the beam telescope was about 1~$\mu$m. 

%--------------------------------------------------
% Sec 2.1 Pixel Hybrids
%--------------------------------------------------
\subsection{Pixel Hybrids}

The prototype pixel sensors are so-called ``n-in-n'' devices: they are designed to collect charge from n$^+$ structures implanted into n- bulk silicon. This design is thought to provide greater longevity in high radiation fields and to allow ``partially-depleted'' operation after ``type-inversion'' of the substrate.  It is more expensive than the "p-in-n" process commonly used in strip detectors because it requires double-sided processing and the implementation of inter-pixel isolation. Two isolation techniques were tested: p-spray, where a uniform medium dose of p-impurities covers the whole structured surface, and p-stop, where higher dose rings individually surround the n$^+$ implants.  Results on the Lorentz angle and charge collection efficiency measurements as well as a detailed description of both designs can be found elsewhere \cite{andrei,tilman}.  In this paper we discuss only measurements performed on p-spray sensors.  All test devices were 22$\times$32 arrays of 125$\times$125~$\mu$m$^2$ pixels having a sensitive area of 2.75$\times$4~mm$^2$.  The substrate was 285~$\mu$m thick, n-doped, diffusively-oxygenated silicon of orientation $\langle111\rangle$, resistivity 2-5~k$\Omega\cdot$cm and oxygen concentration in the order of $10^{17}$ cm$^{-3}$.  Individual sensors were diced from fully processed wafers after the deposition of under-bump metalization and indium bumps.  A number of sensors were irradiated at the CERN PS with 24 GeV protons. The irradiation was performed without cooling or bias. The delivered proton fluences scaled to 1 MeV neutrons by the hardness factor 0.62~\cite{hardfactor}
%\footnote{All particle fluences are normalized to 1 MeV neutrons (${\rm n_{eq}}/\mbox{cm}^2$).}
 were $6\times$10$^{14}$~n$_{\rm eq}$/cm$^2$ and $8\times$10$^{14}$~n$_{\rm eq}$/cm$^2$. The former sample was annealed for three days at 30$^\circ$C. In order to avoid reverse annealing, the sensors were stored at -20$^\circ$C after irradiation and kept at room temperature only for transport and bump bonding. All sensors were bump bonded to PSI30/AC30 readout chips \cite{psi30} which allow analog readout of all 704 pixel cells without zero suppression.  The PSI30 also has a linear response to input signals ranging from zero to more than 30,000 electrons.

%--------------------------------------------------
% Sec 2.2 Data
%--------------------------------------------------
\subsection{Data}

\begin{figure}[hbt]
%
% Figure 2
%
  \begin{center}
    \resizebox{\linewidth}{!}{\includegraphics{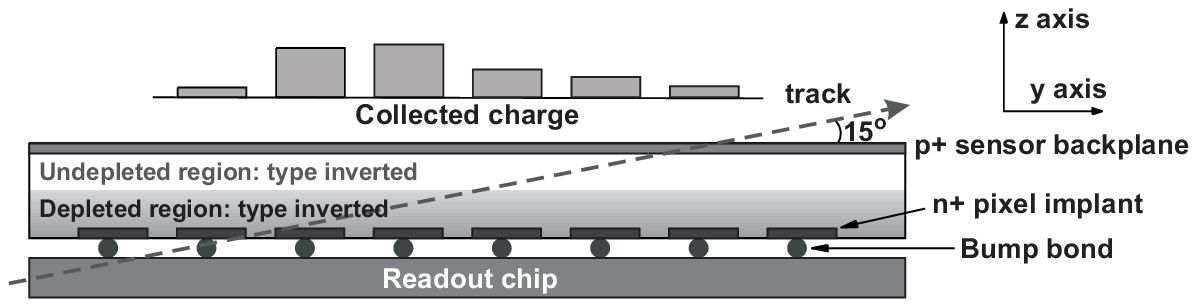}}
  \caption{The grazing angle technique for determining charge collection profiles.  The cluster length is proportional to the depth over which charge is collected.}
  \label{fig:fifteen_deg} 
  \end{center}
\end{figure}

The main focus of the work presented in this paper involves a set of charge collection measurements that were performed using the ``grazing angle technique'' \cite{grazing_angle}.  As is shown in Fig.~\ref{fig:fifteen_deg}, the surface of the test sensor is oriented by a small angle (15$^\circ$) with respect to the hadron beam.  A large sample of data is collected with zero magnetic field and at a temperature of $-10^\circ$C.  The charge measured by each pixel along the $y$ direction samples a different depth $z$ in the sensor.  Precise entry point information from the beam telescope is used to produce finely binned charge collection profiles.  For unirradiated sensors, the cluster length determines the depth over which charge is collected in the sensor.

\begin{figure}[hbt]
  \begin{center}
    \resizebox{0.8\linewidth}{!}{\includegraphics{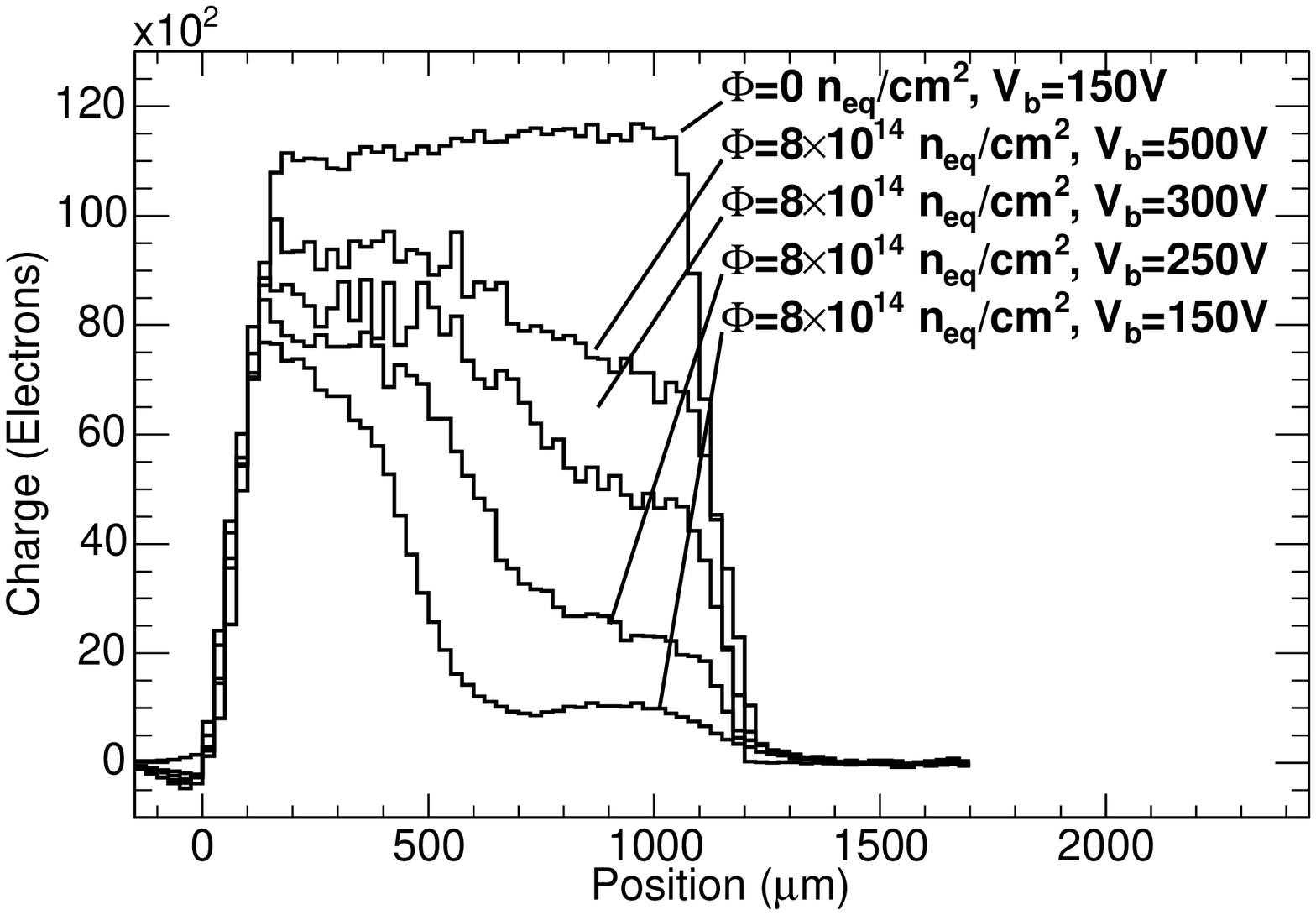}}
  \caption{Charge collection profiles for an irradiated ($\Phi=8\times10^{14}$~n$_{\rm eq}/{\rm cm^2}$) and an unirradiated sensor ($\Phi=0$~n$_{\rm eq}/{\rm cm^2}$) operated at several bias voltages.}
  \label{fig:data} 
  \end{center}
\end{figure}

The profiles that were observed for an unirradiated sensor and for a sensor that was irradiated to a fluence of $\Phi=8\times10^{14}$~n$_{\rm eq}/{\rm cm}^{2}$ are shown in Fig.~\ref{fig:data} as function of the distance from the beam entry point.  The unirradiated sensor was operated at a bias voltage of 150~V which is well above its depletion voltage (approximately 70~V).  It produces a large and uniform collected charge distribution indicating that it is fully depleted (a large $y$ coordinate indicates a large collection distance).  The irradiated sensor was operated at  bias voltages varying from 150~V to 600~V.  It appears to be partly depleted at 150~V, however, signal is collected across the entire thickness of the detector.  Another puzzle is that the ratio of the charges collected at 300~V bias and 150~V bias and integrated along the distance to the beamentry point is 2.1 which is much larger than the maximum value of $\sqrt{2}$ expected for a partially depleted junction, where the depletion depth is proportional to the square root of the bias voltage~\cite{sze}.  It is clear that the profiles for the irradiated sensor exhibit rather different behavior than one would expect for a heavily-doped, unirradiated sensor.

%-------------------------------------------------------
% Sec 3. Simulation and comparison with data
%-------------------------------------------------------
\section{Simulation and comparison with data}
\label{sec:sim}
It is well-known that carrier trapping is a significant effect in heavily irradiated silicon detectors.  In order to evaluate the effects of trapping, it is necessary to implement a detailed simulation of the sensor.  Our simulation, PIXELAV \cite{pixelav1,pixelav2}, incorporates the following elements: an accurate model of charge deposition by primary hadronic tracks (in particular to model delta rays); a realistic electric field map resulting from the simultaneous solution of Poisson's Equation, continuity equations, and various charge transport models; an established model of charge drift physics including mobilities, Hall Effect, and 3-D diffusion; a simulation of charge trapping and the signal induced from trapped charge; and a simulation of electronic noise, response, and threshold effects.  A final step reformats the simulated data into test beam format so that it can be processed by the test beam analysis software.

Several of the PIXELAV details described in~\cite{pixelav1,pixelav2} have changed since they were published.  The commercial semiconductor simulation code now used to generate a full three dimensional electric field map is the ISE TCAD package \cite{ise}.  The charge transport simulation was modified to integrate only the fully-saturated drift velocity,
\begin{equation}
\frac{d\vec r}{dt} = \frac{\mu\left[q\vec E +\mu R_H\vec E\times\vec B+q\mu^2 R_H^2(\vec E\cdot\vec B)\vec B\right ]}{1+\mu^2 R_H^2 |\vec B|^2}
\label{eq:veq}
\end{equation}
where $\mu(\vec{E})$ is the mobility, $q=\pm1$ is the sign of the charge carrier, $\vec E$ is the electric field, $\vec B$ is the magnetic field, and $R_H$ is the Hall factor of the carrier.  The use of the fully-saturated drift velocity permits much larger integration steps (which had previously been limited by stability considerations) and significantly increases the speed of the code.

\begin{figure}[hbt]
%
% Figure 4
%
  \begin{center}    
    \mbox{
      \subfigure[]{\scalebox{0.48}{
	  \epsfig{file=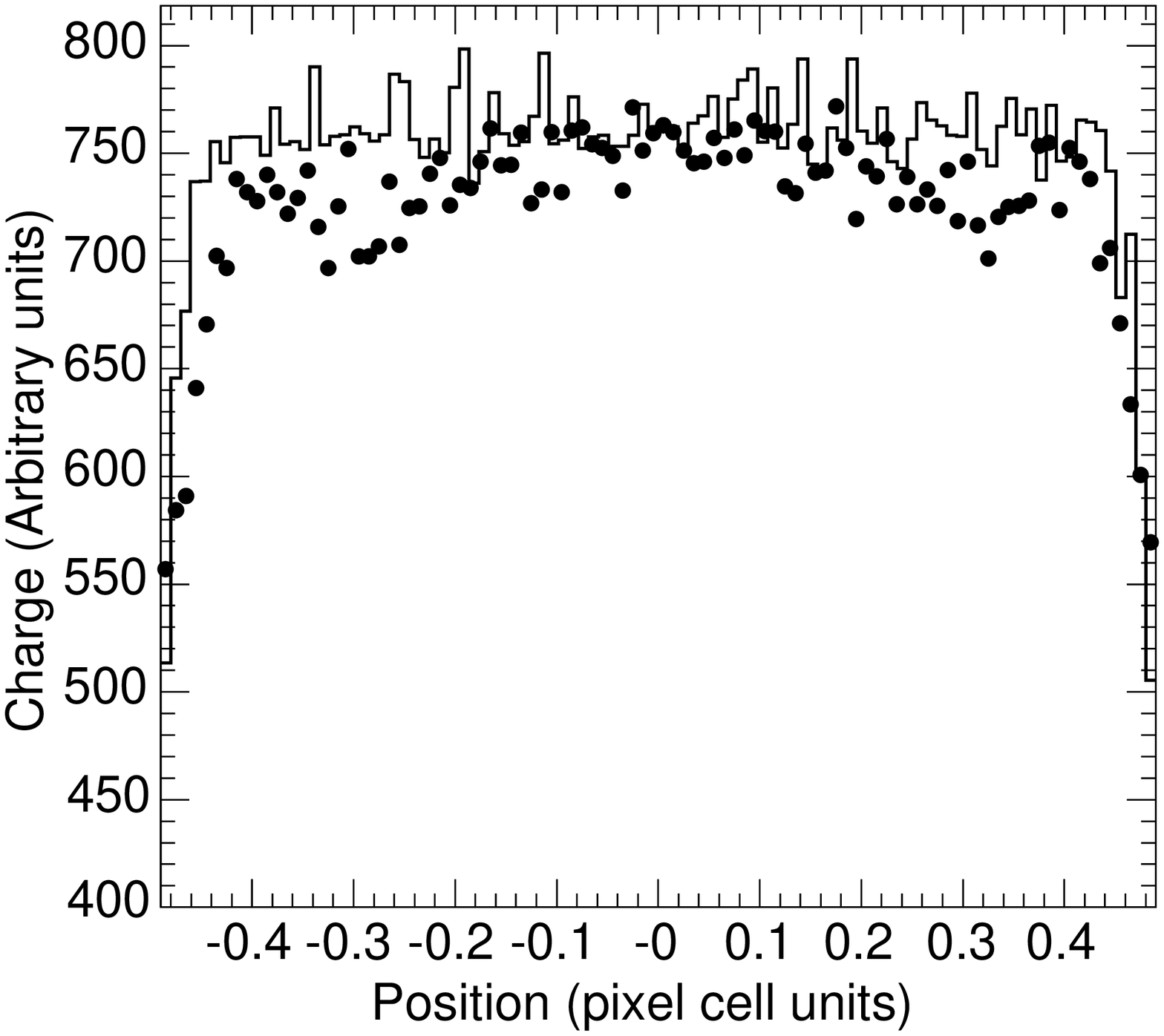,width=\linewidth}
	  \label{fig:sharing_x}
      }}
      \subfigure[]{\scalebox{0.48}{
	  \epsfig{file=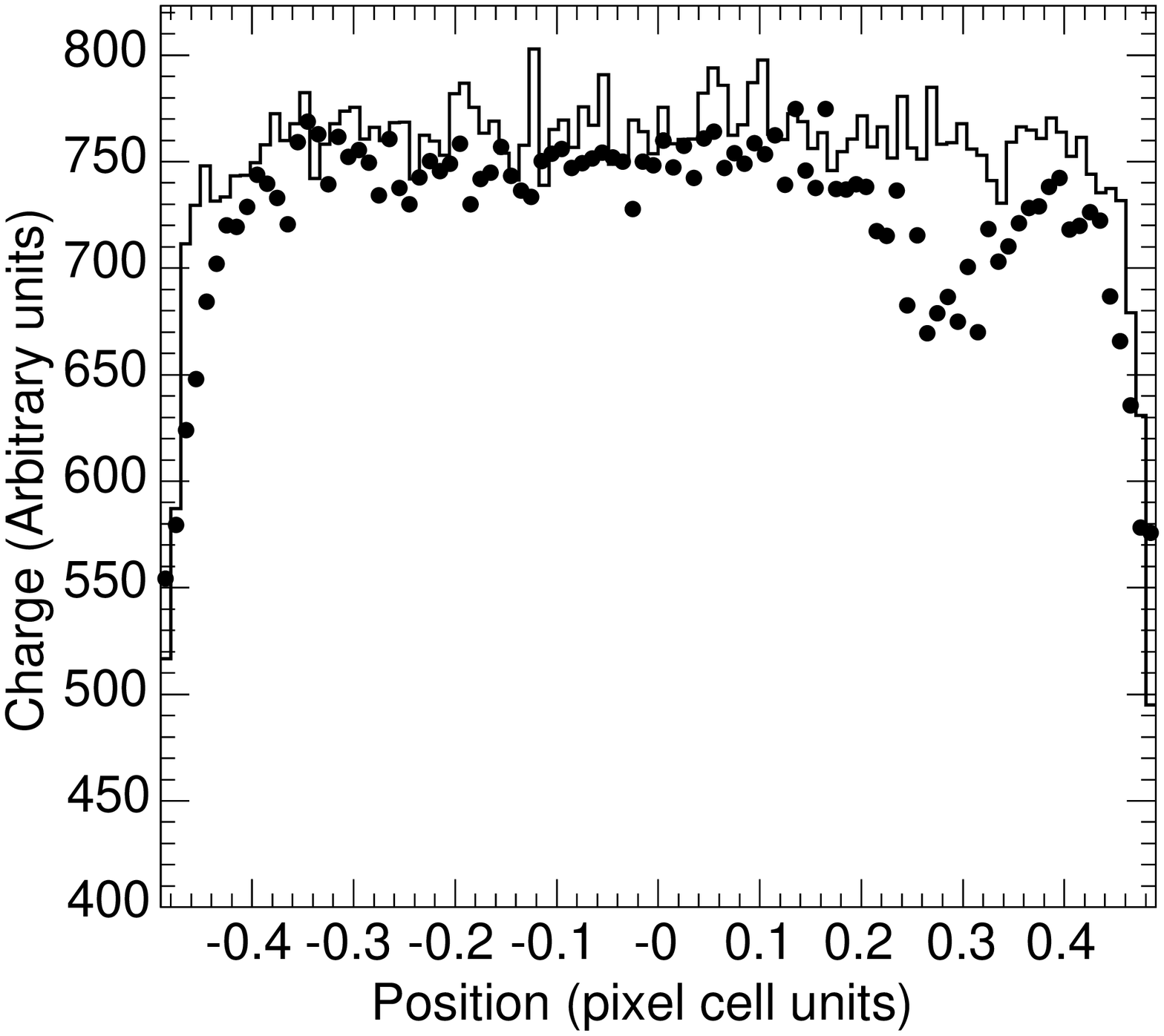,width=\linewidth}
	  \label{fig:sharing_y}
      }}
    }
    %\resizebox{\linewidth}{!}{\includegraphics{compare_intrapixel.eps}}
  \caption{Collected charge measured in a single pixel as a function of the horizontal (a) and vertical (b) track impact position for tracks that are normally incident on an unirradiated sensor.  The simulation is shown as a solid histogram and the test beam data are shown as solid dots.\label{fig:sharing}}
  \end{center}
\end{figure}

The simulation was checked by comparing simulated data with measured data from an unirradiated sensor. 
%The results are summarized in Ref.~\cite{physics0409049}.  
A plot of the charge measured in a single pixel as a function of the horizontal and vertical track impact position for normally incident tracks is shown in Fig.~\ref{fig:sharing}.  The simulation is shown as the solid histogram and the test beam data are shown as solid points.  Note that the sensor simulation does not include the ``punch-through'' structure on the n$^+$ implants which is used to provide a high resistance connection to ground and to provide the possibility of on-wafer IV measurements~\cite{atlas}.  There is reduced charge collection from this portion of the implant and the data show reduced signal in both projections at the bias dot.  Another check, shown in Table~\ref{tab:lorentz}, is the comparison of the average Lorentz angle measured at several bias voltages \cite{andrei}.  In both cases, reasonable agreement is observed between measured and simulated data.

\begin{figure}[hbt]
%
% Figure 5
%
  \begin{center}
    \mbox{
      \subfigure[]{\scalebox{0.48}{
	  \epsfig{file=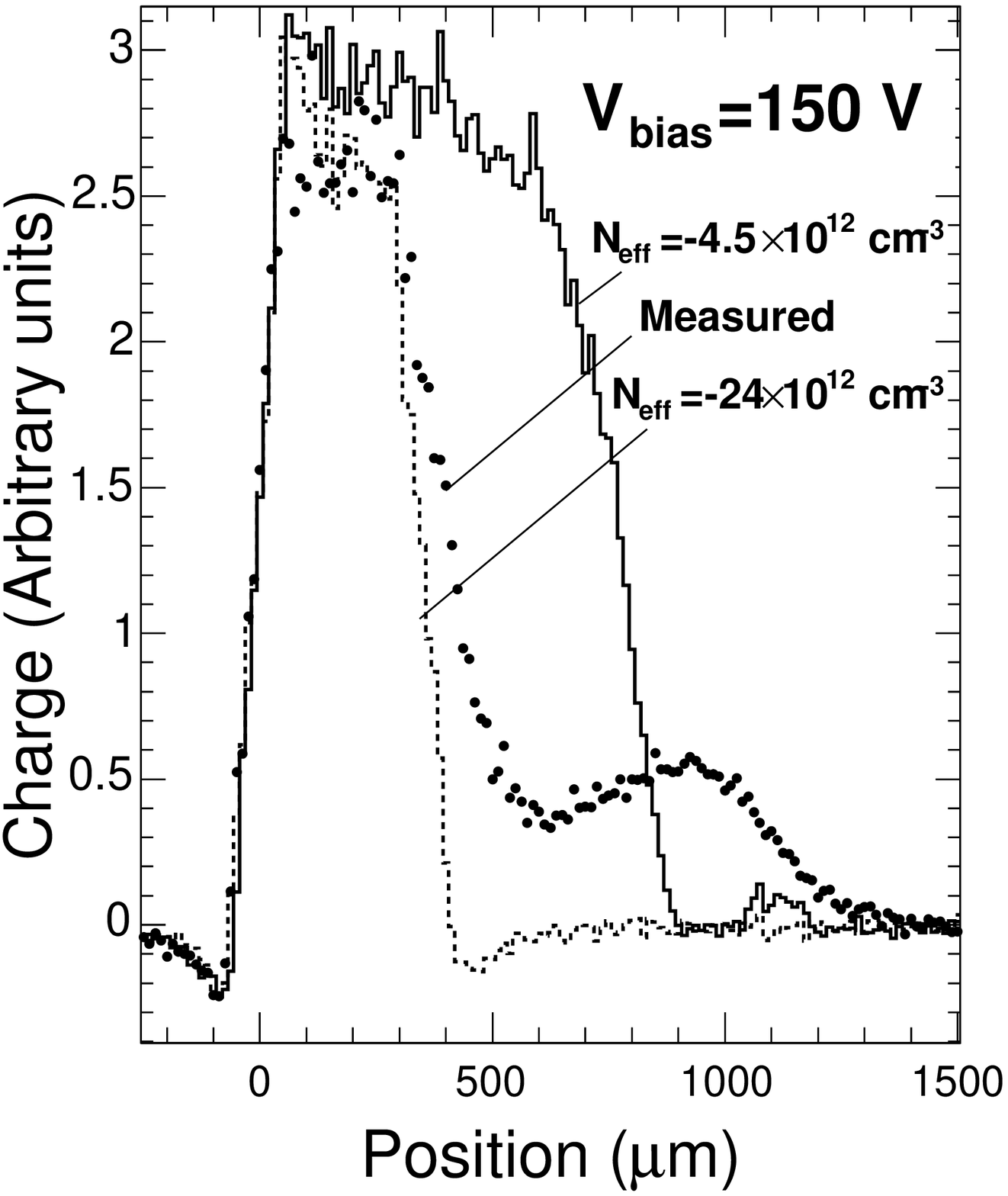,width=\linewidth}
	  \label{fig:strawman_a}
      }}
      \subfigure[]{\scalebox{0.48}{
	  \epsfig{file=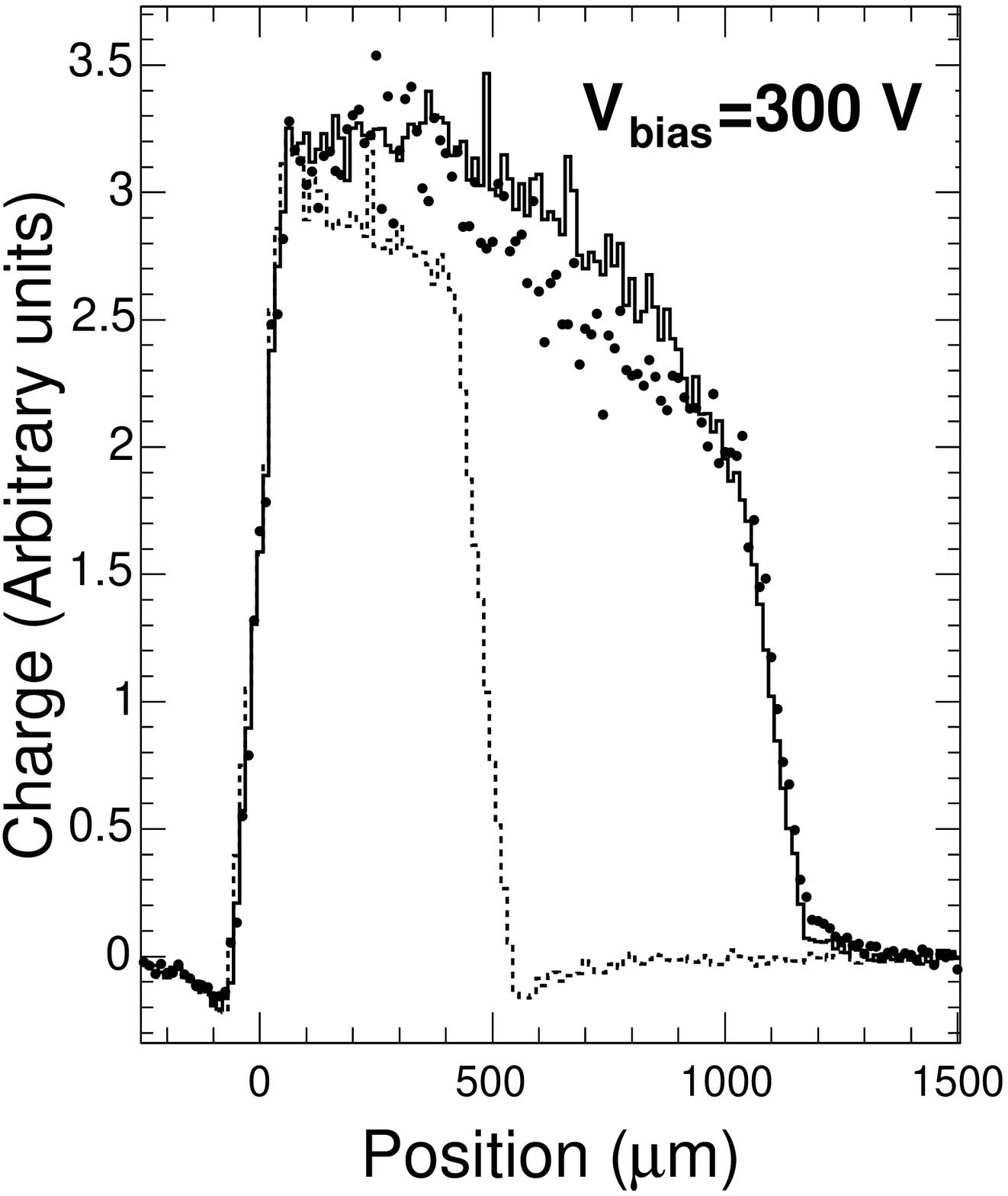,width=\linewidth}
	  \label{fig:strawman_b}
      }}
    }
    %\resizebox{\linewidth}{!}{\includegraphics{IEEE_strawman.eps}}
  \caption{The measured and simulated charge collection profiles for a sensor irradiated to a fluence of $\Phi=6\times10^{14}$~n$_{\rm eq}/$cm$^{2}$.  The profiles measured at bias voltages of 150 V (a) and 300 V (b) are shown as solid dots.  The full histograms are the simulated profiles for a constant effective doping  $N_{\rm eff}=-4.5\times10^{12}$~cm$^{-3}$ of acceptor impurities.  The dashed histograms are the simulated profiles for a constant effective doping  $N_{\rm eff}=-24\times10^{12}$~cm$^{-3}$.} 
  \end{center}
\end{figure}

\begin{table}[h]
\caption{Measured and simulated values of average Lorentz angle $\theta_L$ versus bias voltage for an unirradiated sensor}
\begin{center}
\begin{tabular}{ccc}
\hline
Bias Voltage [V] & Measured $\theta_L$ [deg]& Simulated $\theta_L$ [deg] \\
\hline
150 & $22.8\pm0.7$ & 24.7$\pm$0.9 \\
300 & $14.7\pm0.5$ & 17.4$\pm$0.9 \\
450 & $11.2\pm0.5$ & 12.0$\pm$0.9  \\
\hline
\end{tabular}
\end{center}
\label{tab:lorentz}
\end{table}
The charge collection profiles for a sensor irradiated to a fluence of $\Phi=6\times10^{14}$~n$_{\rm eq}/{\rm cm}^{2}$ and operated at bias voltages of 150~V and 300~V are presented in Fig.~\ref{fig:strawman_a} and~\ref{fig:strawman_b}, respectively.  The measured profiles are shown as solid dots and the simulated profiles are shown as histograms.  The simulated profiles were generated with electric field maps corresponding to two different effective densities of acceptor impurities.  The full histograms are the simulated profile for $N_{\rm eff}=-4.5\times10^{12}$~cm$^{-3}$.  Note that the 300 V simulation reasonably agrees with the measured profile but the 150 V simulation is far too broad.  The dashed histograms show the result of decreasing $N_{\rm eff}$ to $-24\times10^{12}$~cm$^{-3}$.  At this effective doping density, the width of the simulated peak in the 150V distribution is close to correct but it does not reproduce the second peak observed in the data at large $y$.  The 300 V simulated distribution is far too narrow and the predicted charge is lower than the data (note that the profiles are absolutely normalized).  It is clear that a simulation based upon the standard picture of a constant density of ionized acceptor impurities cannot reproduce the measured profiles.

Note that the simulation of this irradiated sensor includes the effects of trapping.  The trapping rates of electron and holes have been shown to scale linearly with fluence \cite{kramberger1}.  Unfortunately, the measured fluences have a fractional uncertainty of $\pm$10\% which feeds directly into the uncertainty on the trapping rates.  Additional uncertainty arises because annealing can modify the trapping rates by 30\% \cite{kramberger1} leading to a fairly large overall uncertainty.  
%The charge collection profiles are not very sensitive to the hole trapping rates but are sensitive to the electron trapping rates. 
%The profiles shown in Fig.~\ref{fig:strawmen} are simulated by reducing the trapping rates by 20\% with respect to the nominal rates for reasons that will be explained in Section~\ref{sec:djmodels}.  The choice does not affect the conclusion that the measured profiles are inconsistent with a constant effective doping model.

%---------------------------------------------------------
% Two-trap Models
%---------------------------------------------------------
\section{Two-trap Models}
\label{sec:djmodels}

\begin{figure}[hbt]
%
% Figure 6
%
  \begin{center}
    \resizebox{\linewidth}{!}{\includegraphics{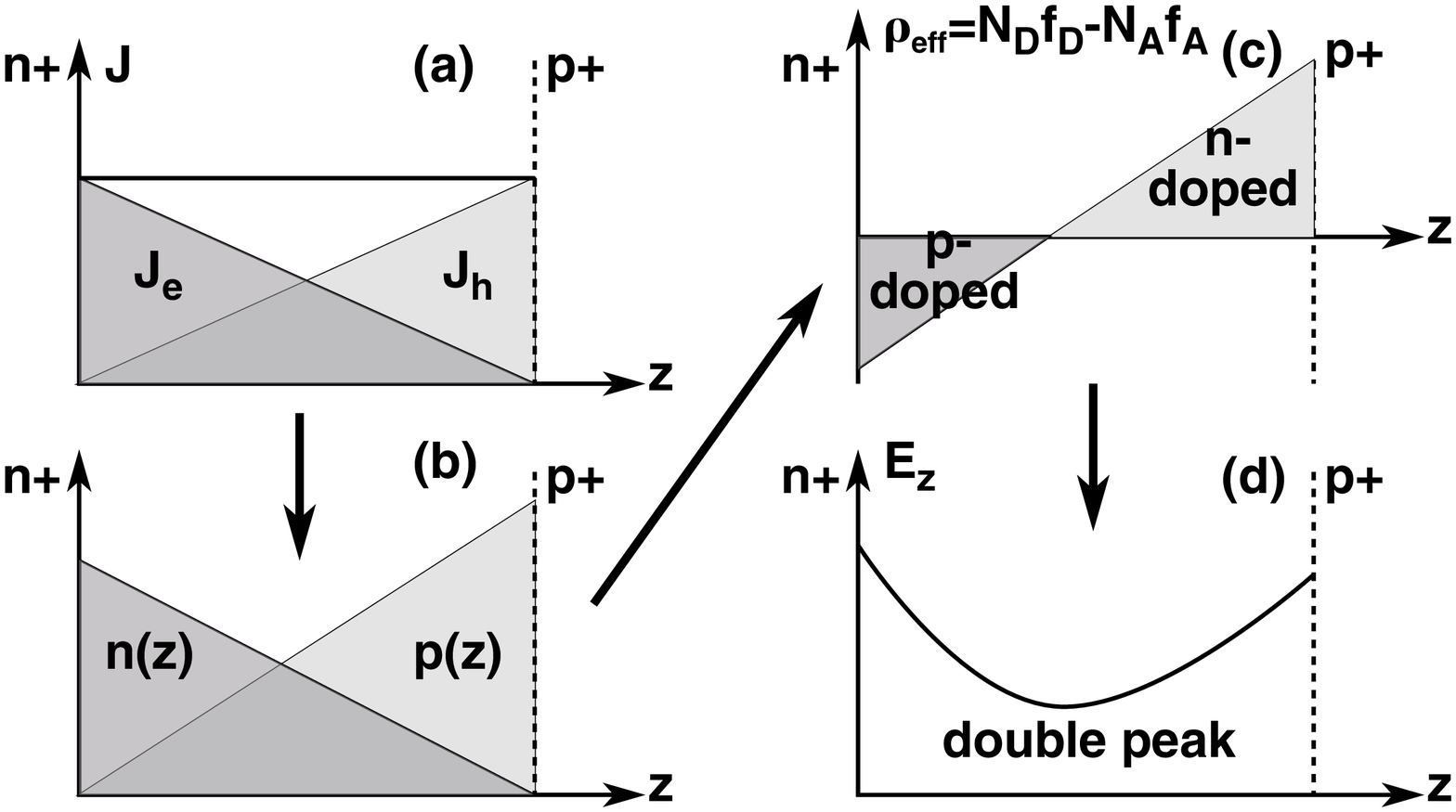}}
  \caption{An illustrative sketch of the EVL model for a reverse biased device~\cite{evl}.}
  \label{fig:evl} 
  \end{center}
\end{figure}

The large number of measurements suggesting that large electric fields exist at both sides of an irradiated silicon diode has given rise to several attempts to model the effect \cite{cgg,castaldini,evl}.  The most recent of these by Eremin, Verbitskaya, and Li (EVL) \cite{evl} is based upon a modification of the Shockley-Read-Hall (SRH) statistics.  The EVL model produces an effective space charge density $\rho_\mathrm{eff}$ from the trapping of free carriers originated from the leakage current by one acceptor trap and one donor trap.  
The effective charge density is related to the occupancies and densities of traps as follows,
\begin{equation}
\rho_\mathrm{eff} = e\left[N_Df_D-N_Af_A\right] + \rho_\mathrm{dopants}   \label{eq:rhodef}
\end{equation}
where: $N_D$ and $N_A$ are the densities of donor and acceptor trapping states, respectively; $f_D$ and $f_A$ are the occupied fractions of the donor and acceptor states, respectively, and $\rho_\mathrm{dopants}$ is the charge density due to ionized dopants.  
Charge flows to and from the trapping states due to generation and recombination.  The occupied fractions are given by the following standard SRH expressions,
\begin{eqnarray}
f_D &=& \frac{v_h\sigma^D_hp+v_e\sigma^D_en_ie^{E_D/kT}}{v_e\sigma^D_e(n+n_ie^{E_D/kT})+v_h\sigma^D_h(p+n_ie^{-E_D/kT})}    \label{eq:fDAdef}  \\
f_A &=& \frac{v_e\sigma^A_en+v_h\sigma^A_hn_ie^{-E_A/kT}}{v_e\sigma^A_e(n+n_ie^{E_A/kT})+v_h\sigma^A_h(p+n_ie^{-E_A/kT})}  \nonumber
\end{eqnarray}
where: $v_e$ and $v_h$ are the thermal velocities of electrons and holes, respectively; $\sigma_e^D$ and $\sigma_h^D$ are the electron and hole capture cross sections for the donor trap; $\sigma_e^A$ and $\sigma_h^A$ are the electron and hole capture cross sections for the acceptor trap; $n$ and $p$ are the densities of free electrons and holes, respectively; $n_i$ is the intrinsic density of carriers; $E_D$ and $E_A$ are the activation energies (relative to the mid-gap energy) of the donor and acceptor states, respectively.  The generation-recombination current caused by the SRH statistics for single donor and acceptor states is given by the following expression,
\begin{eqnarray}
U &=& \frac{v_hv_e\sigma^D_h\sigma^D_eN_D(np-n_i^2)}{v_e\sigma^D_e(n+n_ie^{E_D/kT})+v_h\sigma^D_h(p+n_ie^{-E_D/kT})} \nonumber \\
&+& \frac{v_hv_e\sigma^A_h\sigma^A_eN_A(np-n_i^2)}{v_e\sigma^A_e(n+n_ie^{E_A/kT})+v_h\sigma^A_h(p+n_ie^{-E_A/kT})}.  \label{eq:Udef}
\end{eqnarray}
Within the EVL model, the four trapping cross sections are set to $10^{-15}$~cm$^2$.  The leakage current is generated from an additional SRH trapping state that is introduced for this purpose but is assumed not to trap charge.  The donor and acceptor states are assumed not to generate leakage current which, given the small size of the cross sections, is a self-consistent assumption. The densities of the donor and acceptor states ($N_D$ and $N_A$) are adjusted to ``fit'' data obtained with the Transient Current Technique (TCT).  The parameters of the model are given in Table~\ref{tab:evl}.  The trap densities are scaled to fluence and are given in terms of introduction rates $g_\mathrm{int}=N_{A/D}/\Phi_\mathrm{eq}$.
\begin{table}[h]
\caption{Parameters of the EVL Model \cite{evl}. $E_V$ and $E_C$ are the valence and conduction band energy levels, respectively}
\begin{center}
\begin{tabular}{ccccc}
\hline
Trap & E (eV) & $g_\mathrm{int}$ (cm$^{-1}$) & $\sigma_e$ (cm$^2$) & $\sigma_h$ (cm$^2$)  \\
\hline
Donor    & $E_V+0.48$  & 6   & $1\times10^{-15}$ & $1\times10^{-15}$ \\
Acceptor & $E_C-0.525$ & 3.7 & $1\times10^{-15}$ & $1\times10^{-15}$ \\
\hline
\end{tabular}
\end{center}
\label{tab:evl}
\end{table}
An illustrative sketch of the EVL model has been reproduced from~\cite{evl} and is shown in Fig.~\ref{fig:evl}.  Figure~\ref{fig:evl}(a) shows a uniform current density flowing across a reverse-biased junction.  Since holes are produced uniformly across the junction and flow to the p$^+$ backplane, the hole current density increases linearly with increasing $z$ from the n$^+$ implant to the p$^+$ implant.  The electrons flow to the n$^+$ implant and the electron current density increases with decreasing $z$.  The actual carrier densities depend upon the details of the fields and mobilities but vary monotonically across the junctions as shown in Fig.~\ref{fig:evl}(b).  The trapping of the mobile carriers produces a net positive space charge density near the p$^+$ backplane and a net negative space charge density near the n$^+$ implant as shown in Fig.~\ref{fig:evl}(c).  Since positive space charge corresponds to n-type doping and negative space charge corresponds to p-type doping, there are p-n junctions at both sides of the detector.  The electric field in the sensor follows from a simultaneous solution of Poisson's equation and the continuity equations.  The resulting $z$-component of the electric field is shown in Fig.~\ref{fig:evl}(d).  It varies with an approximately quadratic dependence upon $z$ having a minimum at the zero of the space charge density and maxima at both implants.  

In order to test whether the electric field predicted by the EVL model would improve the agreement between the simulated and measured charge collection profiles shown in Section~\ref{sec:sim}, it was necessary to implement the EVL model in ISE TCAD.  TCAD contains a complete implementation of SRH statistics.  However, the EVL modifications of SRH are not incorporated.  In particular, any state added to generate leakage current would also trap charge.  It is possible to use the TCAD ``Physical Model'' interface to replace the existing SRH implementation with a modified one, however, a less invasive approach was adopted.

Our approach is based upon three observations:
\begin{enumerate}
\item
The trapping cross sections are poorly known.  The cross sections for states observed in various types of defect spectroscopy vary over several orders of magnitude.
\item
The occupancies $f_{D/A}$ of the trapping states are independent of the scale of the cross sections.  If the capture cross sections $\sigma_{e/h}$ in Eq.~\ref{eq:fDAdef} are rescaled by a factor $r$, then $f_{D/A}$ are unchanged.  The occupancies depend only upon the ratio of electron and hole cross sections, $\sigma_h/\sigma_e$.
\item
The current $U$ generated by the donor and acceptor impurities is linear in the cross section rescaling factor $r$.
\end{enumerate}
These observations imply that it is possible to implement the EVL model in TCAD simply by setting $\sigma^D_e=\sigma^D_h=\sigma^A_e=\sigma^A_h$ and by varying the size of the common cross section until the generation current is equal to the observed or expected leakage current.  The trap occupancies are not affected in zeroth order by the rescaling, but the leakage current and the free carrier densities are affected by $r$.  The carrier densities have a first-order effect on the occupancies so that varying $r$ does alter $\rho_\mathrm{eff}$.  This approach uses the same trapping states to produce space charge and leakage current (it is not necessary to introduce current-generating states).  

Using the activation energies and introduction rates for the donor and acceptor states given in Table~\ref{tab:evl}, the only free parameter in the TCAD implementation of the EVL model is the size of the common cross section or equivalently, the leakage current.  The leakage current in the test sensors was observed to increase substantially after bump-bonding to the readout chips.  Although the cause of the increased current is not clear, it is possible that it is due to increased surface/edge leakage or that it is caused by stressing the detector.  
The leakage current of the sample irradiated to $6\times10^{14}$~n$_{\rm eq}/$cm$^{2}$ at a bias voltage of 300 V and temperature of -10$^\circ$C was 15.1 $\mu$A. 
The expected value calculated using Eq.~1 and 2 of~\cite{mfl} and a damage rate $\alpha_0= 4\times 10^{-17}$ A/cm is found to be 4.7 $\mu$A.
%\footnote{The expected leakage current was calculated using Eq.~1 and 2 of~\cite{mfl}.}.
%The measured leakage current, expressed in terms of the damage parameter $\alpha$ \cite{mfl}, is listed in Table~\ref{tab:ileak}.  It is well established that in large detectors, the magnitude of leakage current is expected to be $\alpha_0\simeq4\times10^{-17}$~A/cm for a fully biased detector~\cite{mfl}.  The observed leakage current is approximately five times larger than the expected value.
%\begin{table}[h]
%\caption{Observed leakage current in terms of $\alpha=I_\mathrm{leak}(20^\circ\mathrm{C})/(\mathcal{V}\Phi)$ where: $I_\mathrm{leak}(20^\circ C)$ is the leakage current expected at temperature $20^\circ$C, $\mathcal{V}$ is the volume of the sensor, and $\Phi$ is the neutron-equivalent fluence}
%\begin{center}
%\begin{tabular}{cc}
%\hline
%V$_{\rm{bias}}$ [V] & $\alpha$ [A/cm]\\
%\hline
%150 & $15\times10^{-17}$ \\
%300 & $19\times10^{-17}$ \\
%450 & $25\times10^{-17}$ \\
%\hline
%\end{tabular}
%\end{center}
%\label{tab:ileak}
%\end{table}

\begin{figure*}[hbt]
%
%  Figure 7 
%
  \begin{center}
    \mbox{
      \subfigure[]{\scalebox{0.30}{
	  \epsfig{file=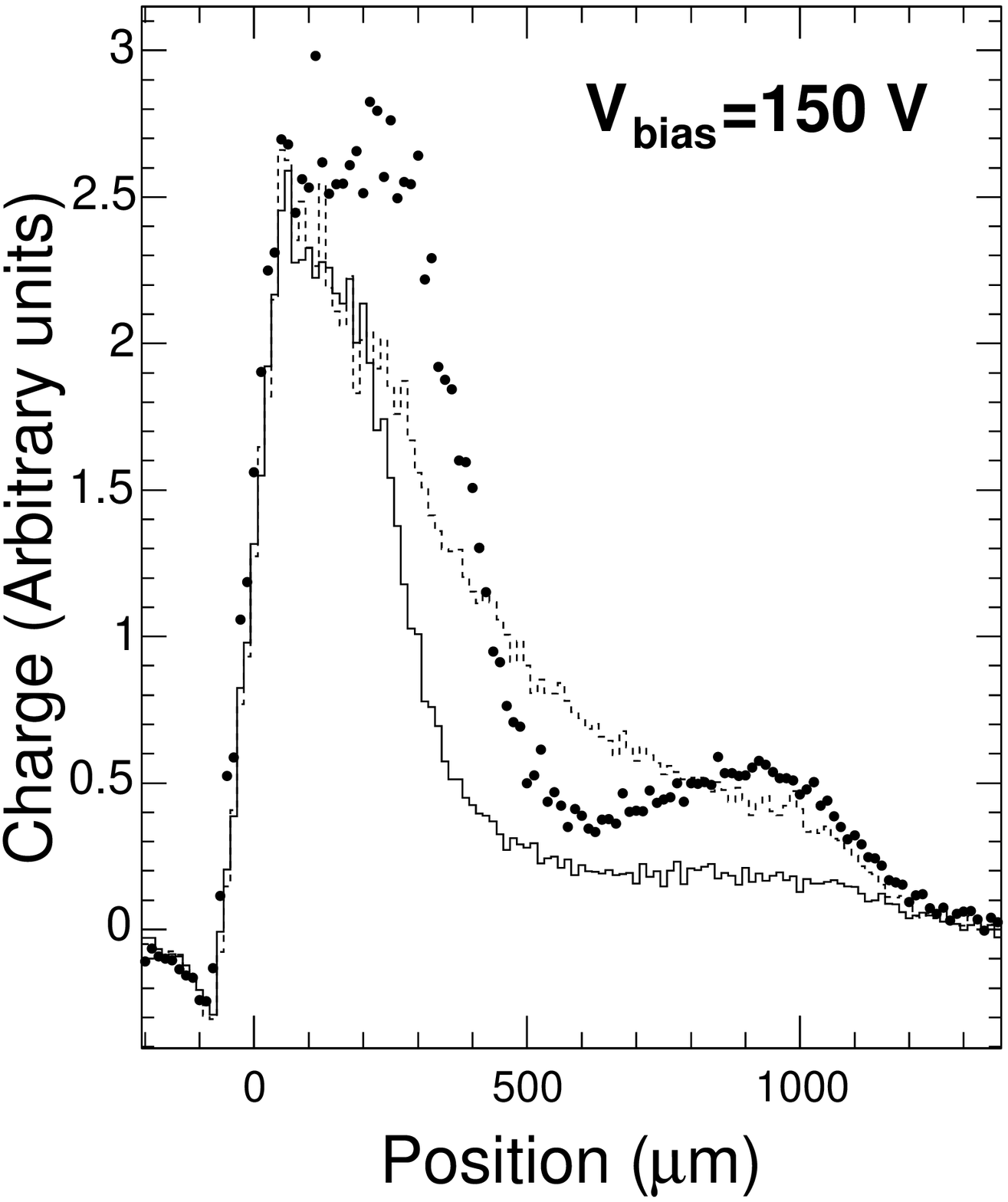,width=\textwidth}
	  \label{fig:evlsim_a}
      }}
      \subfigure[]{\scalebox{0.30}{
	  \epsfig{file=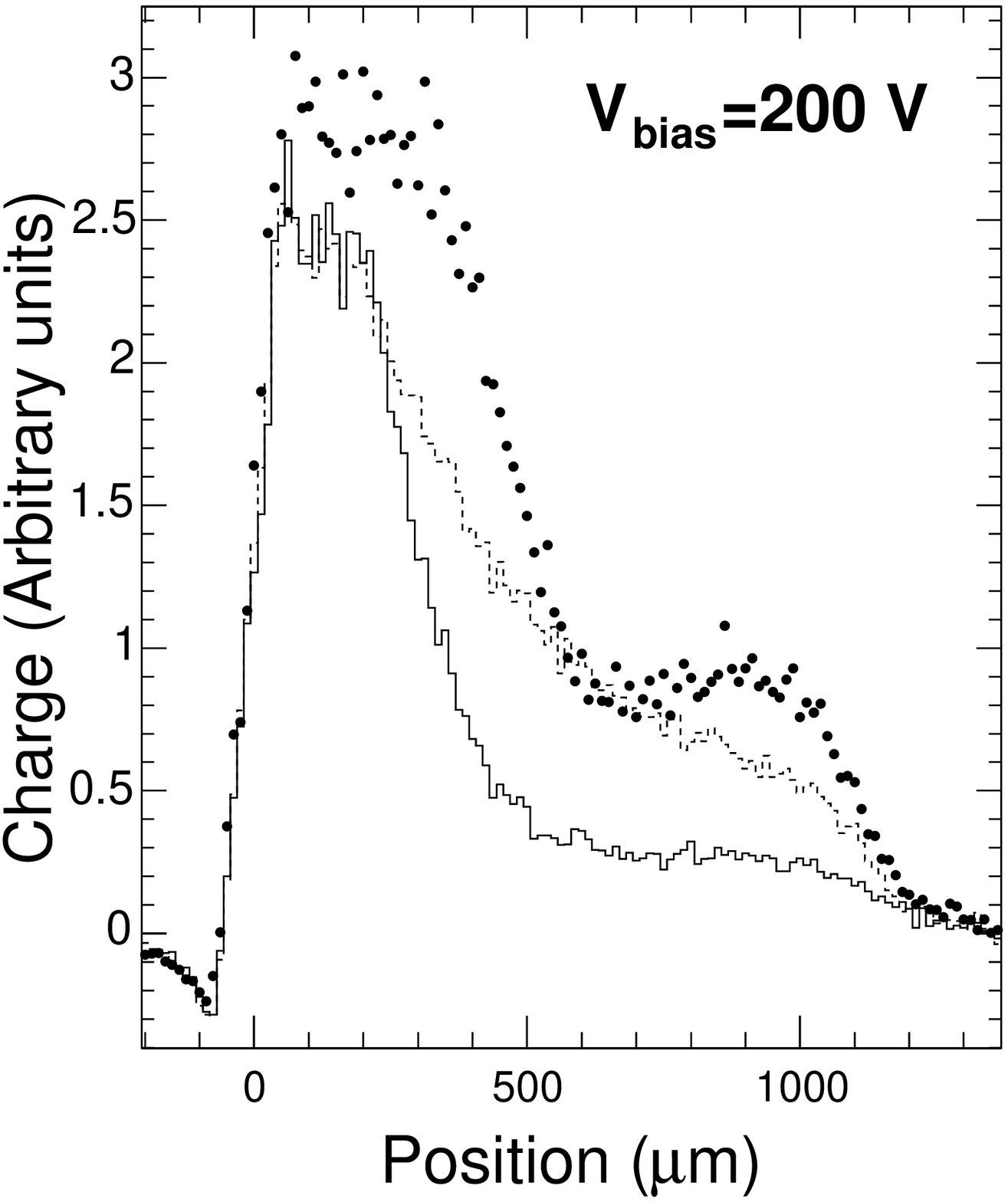,width=\textwidth}
	  \label{fig:evlsim_b}
      }}
      \subfigure[]{\scalebox{0.30}{
	  \epsfig{file=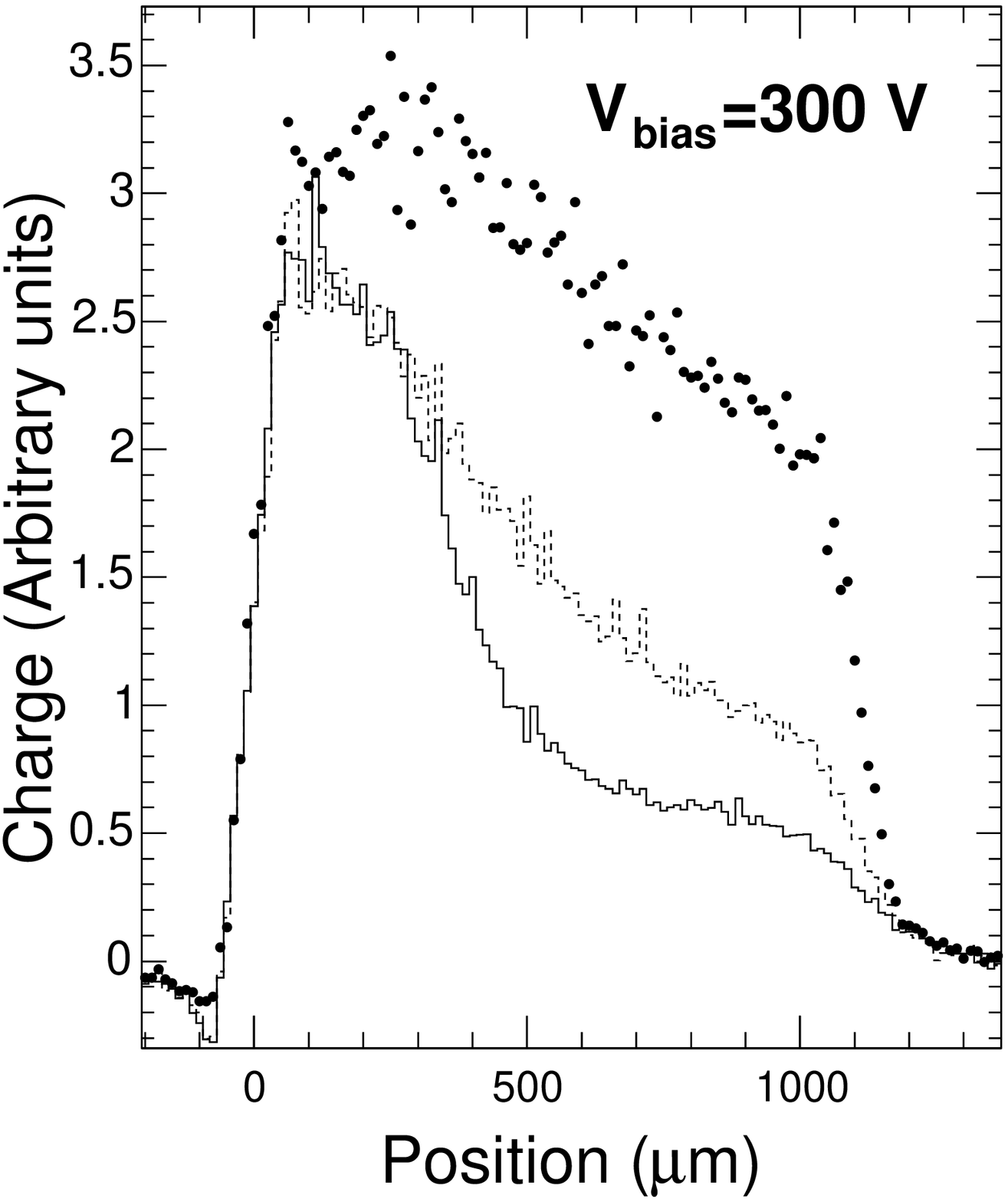,width=\textwidth}
	  \label{fig:evlsim_c}
      }}
    }
    \caption{The measured charge collection profiles for a sensor irradiated to a fluence of $\Phi=6\times10^{14}$~n$_{\rm eq}/$cm$^{2}$ (solid dots) and operated at 150~V (a), 200~V (b) and 300~V (c) are compared with simulations based upon the EVL model.  The solid histogram shows the EVL model with the leakage current adjusted to 
the expected value.  The dashed histogram shows the effect of increasing the leakage current to the measured value. \label{fig:evlsim}}
  \end{center}
\end{figure*}

The EVL model was implemented in the sensor simulation using several parameter choices.  The resulting charge collection profiles for a sensor irradiated to a fluence of $\Phi=6\times10^{14}$~n$_{\rm eq}/$cm$^{2}$ and operated at 150 V, 200~V and 300 V are shown in Fig.~\ref{fig:evlsim}.  The measured profiles are again shown as solid dots.  The solid histogram shows the EVL model with the leakage current adjusted to the expected value by setting $\sigma_{e/h}=0.47\times10^{-14}$ cm$^2$.  It clearly substantially underestimates the collected charge at both voltages.  The effect of increasing the leakage current to the measured value is shown as the dashed histogram, where the trapping cross sections are set to $\sigma_{e/h}=1.48\times10^{-14}$ cm$^2$.  
This increases the signal at 300 V but does not reproduce the shape of the high $z$ tail of the 150 V distribution.  
%Finally, in an attempt to increase the collected charge at 300 V, the introduction rates were scaled down by a factor of eight.  The common cross section was increased by the same factor to hold the leakage current constant.  The result of this is shown as the solid histogram in Fig.~\ref{fig:evlsim_a} and (b).  Note that this does increase the collected charge signal, however, it is still below the data at 300 V and is much too large in the tail region of the 150 V distribution.  
We conclude that the EVL model does not describe the measured charge collection profiles.

%--------------------------------------------------------
% An Improved two-trap model
%--------------------------------------------------------
\section{An Improved Two-Trap Model}
\label{sec:results}

Although the EVL model does not describe the measured charge collection profiles, 
the electric field generated by the two-trap mechanism could potentially reproduce
the main features observed in the data.  
At low bias voltages, the combination of the quadratic minimum in the electric field and 
free carrier trapping can act like a ``gate'' suppressing the collection of charge from the p$^+$ side of the detector.  
The measured profile would then appear to be a narrow peak on the n$^+$ side of the detector.  
As the bias is increased, the magnitude of the field at the minimum would also increase and effectively 
``lift the gate'' which would allow much more charge collection from the p$^+$ side of the detector.

In order to investigate whether a two-trap double junction EVL-like model can describe the measured charge collection profiles, a tuning procedure was adopted.  Relaxing the EVL requirement that all trapping cross sections are equal, the model has six free parameters %(two traps concentrations and four trapping cross sections)
($N_D$, $N_A$, $\sigma_e^D$, $\sigma_h^D$, $\sigma_e^A$, $\sigma_h^A$) 
that can be adjusted.  The activation energies are kept fixed to the EVL values.  Additionally, as was discussed in Section~\ref{sec:sim}, the electron and hole trapping rates, $\Gamma_e$ and $\Gamma_h$, are uncertain at the 30\% level due to the fluence uncertainty and possible annealing of the sensors.  They are treated as constrained parameters.  
%In an ideal case, the simulation would be used to perform a $\chi^2$ fit to the measured profiles and optimal parameter estimates and confidence intervals would be extracted.  Unfortunately, the simulation chain is quite slow.  It requires approximately nine hours to produce a TCAD solution for the electric field map and then several simultaneous Pixelav jobs (one per bias voltage) require approximately 16~hours each to produce adequate statistics.  This implies that a traditional $\chi^2$ fit with a full error matrix analysis is not feasible.  An extremely tedious and less mathematically rigorous procedure was adopted instead.  
The parameters of the double junction model were systematically varied and the agreement between measured and simulated charge collection profiles was judged subjectively.

\begin{figure}[hbt]
  \begin{center}
    \resizebox{\linewidth}{!}{\includegraphics{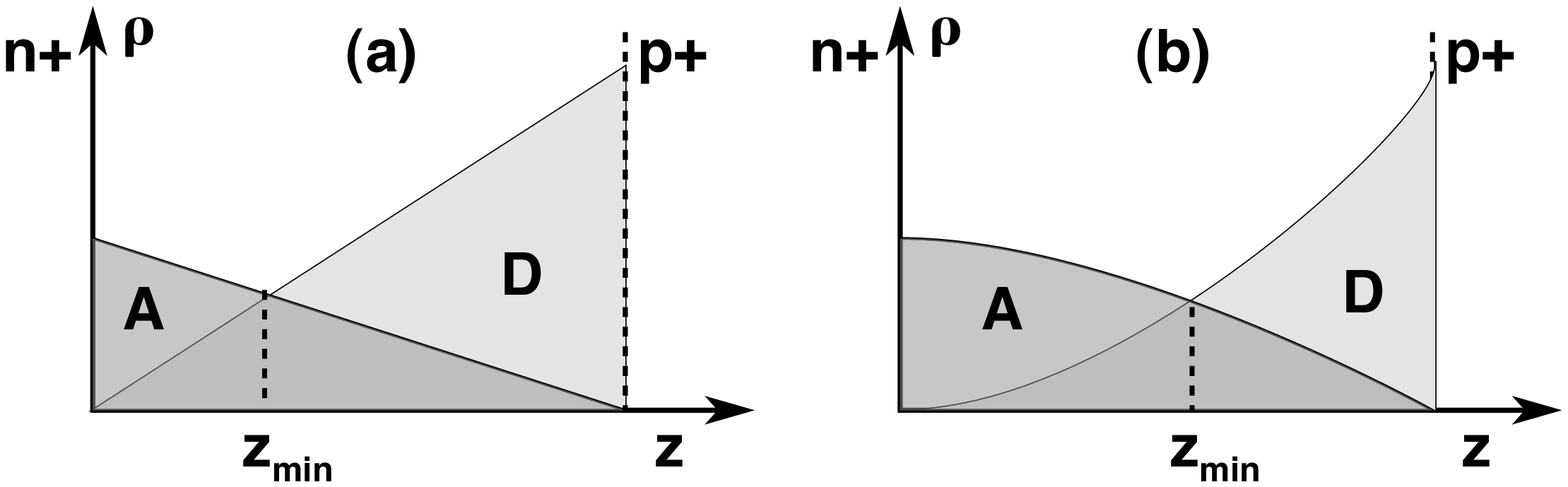}}
  \caption{The effect of increasing $N_D/N_A$ when (a) the electron and hole cross sections are equal, and when (b) $\sigma_h/\sigma_e=0.25$.}
  \label{fig:djcartoon} 
  \end{center}
\end{figure}

In the course of the tuning procedure, it became clear that the EVL model does not produce a sufficiently large electric field on the p$^+$ side of the detector.  The solution to this problem is to increase the density of donors (hole traps) as compared to the density of acceptors (electron traps).  When this is done, the $z$ position of the minimum in the effective charge density shifts toward the n$^+$ implant as sketched in Fig.~\ref{fig:djcartoon}(a).  
Unfortunately, this causes the ``peak'' in the 150 V simulated charge profile to become too narrow.  The position of the charge density minimum can be restored to a position nearer the midplane of the detector by decreasing the ratios of the hole and electron cross sections from 1.0 to 0.25 
($\sigma^D_h/\sigma^D_e=0.25$ and $\sigma^A_h/\sigma^A_e=0.25$) 
as shown schematically in Fig.~\ref{fig:djcartoon}(b).  
Note that the adjustment of the cross sections for both trap types minimizes the field in the quadratic minimum while allowing for large fields at the implants.  For simplicity it is assumed that the electron cross sections are equal ($\sigma^D_e=\sigma^A_e=\sigma_e$)
and that the hole cross sections are equal ($\sigma^D_h=\sigma^A_h=\sigma_h$).

\begin{figure}[hbt]
  \begin{center}
    \resizebox{0.6\linewidth}{!}{\includegraphics[angle=90]{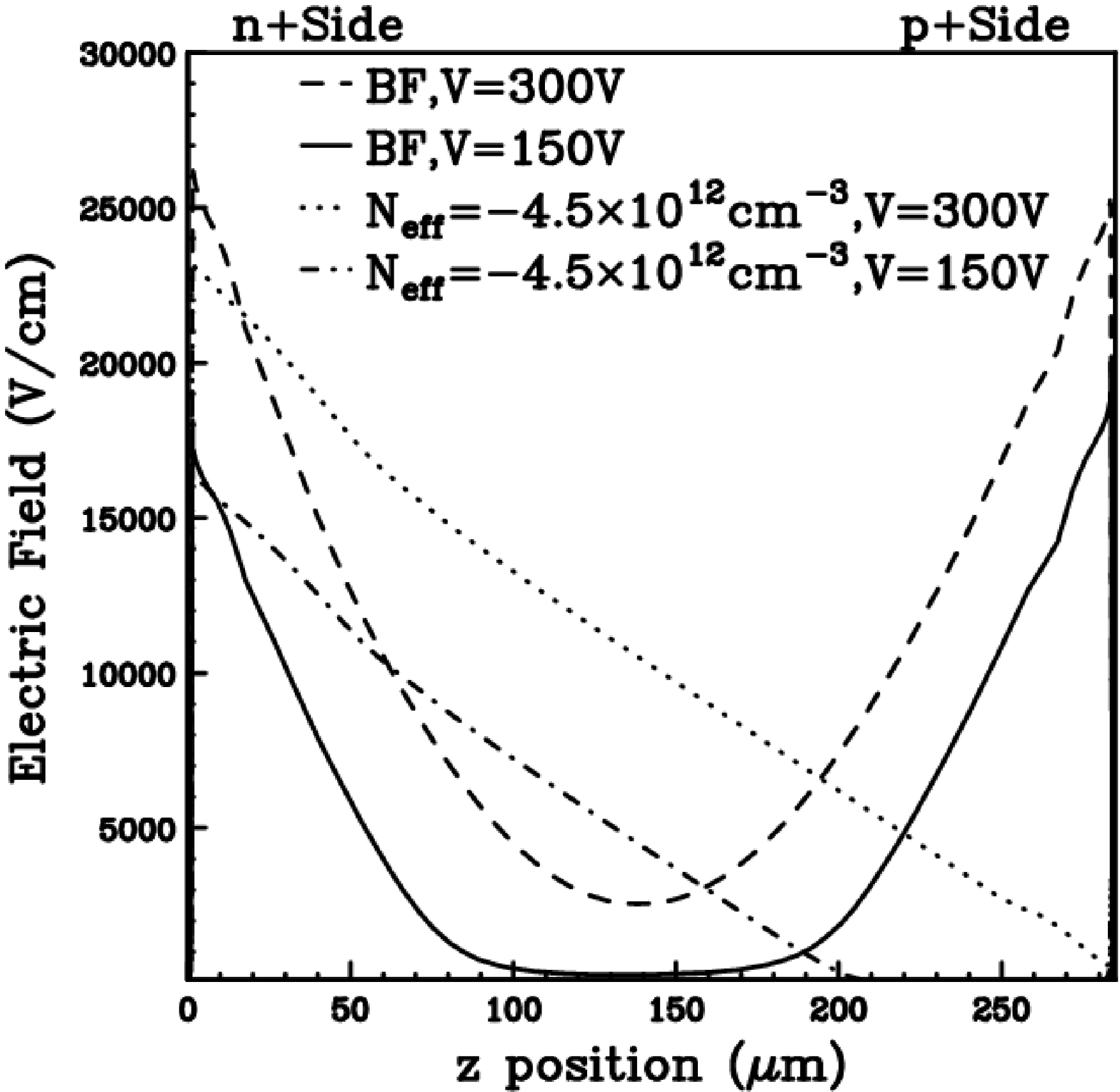}}
  \caption{The $z$-component of the simulated electric field resulting from the model BF is plotted as a function of $z$.  
    The field profiles for 150 V and 300 V are shown as as solid and dashed curves, respectively.  
    %The field profiles that result from the rescaled EVL model are shown as blue solid and dashed curves and labeled as dj16.  
    The electric field profiles resulting from a constant p-type doping of density $N_{\rm eff}=-4.5\times10^{12}$~cm$^{-3}$ 
    are shown as dot-dashed and dotted curves for 150 V and 300 V, respectively.}
  \label{fig:efield} 
  \end{center}
\end{figure}

The current best ``fit'' to the measured charge collection profiles is called 'BF' and reduces the ratio between the densities of acceptor and donor states, $N_A/N_D$,  from the EVL value of 0.62 to 0.40.  The $z$-component of the simulated electric field, $E_z$, is plotted as a function of $z$ in Fig.~\ref{fig:efield} for bias voltages of 150 V and 300 V.  The field profiles have minima near the midplane of the detector.  Note that the minimum field at 150 V bias appears to be very small but is still approximately 400~V/cm.  
%The dj35 field profiles are compared with those that result from the rescaled EVL model which are shown as blue solid and dashed curves (labeled as dj16).  
The electric field profiles resulting from a constant p-type doping of density $N_{\rm eff}=-4.5\times10^{12}$~cm$^{-3}$ are shown as dot-dashed and dotted curves for reference.

\begin{figure*}[!t]
  \centering
  \mbox{
      \subfigure[]{\scalebox{0.24}{
	  \epsfig{file=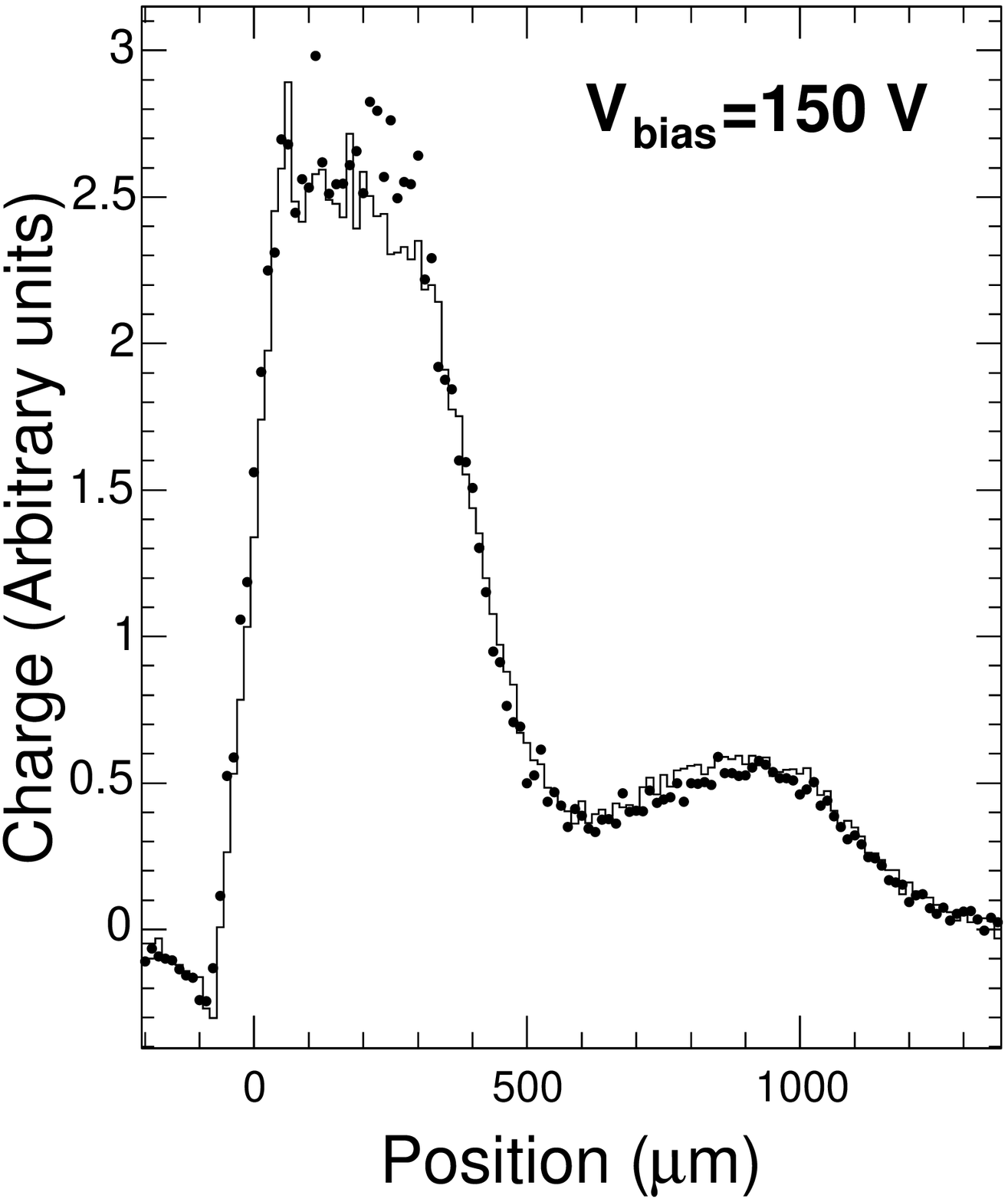,width=\textwidth}
	  \label{fig:dj44_a}
      }}
      \subfigure[]{\scalebox{0.24}{
	  \epsfig{file=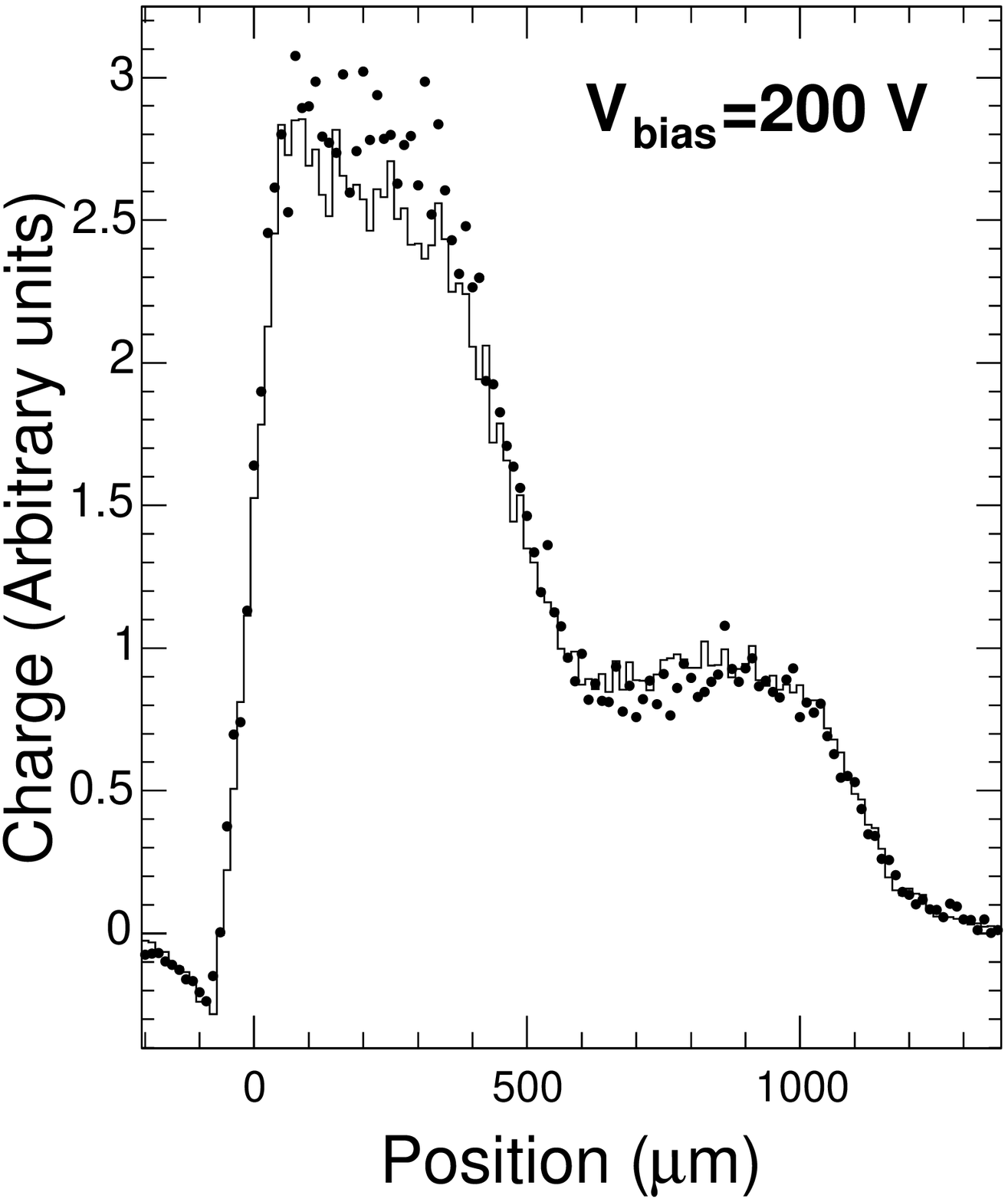,width=\textwidth}
	  \label{fig:dj44_b}
      }}
      \subfigure[]{\scalebox{0.24}{
	  \epsfig{file=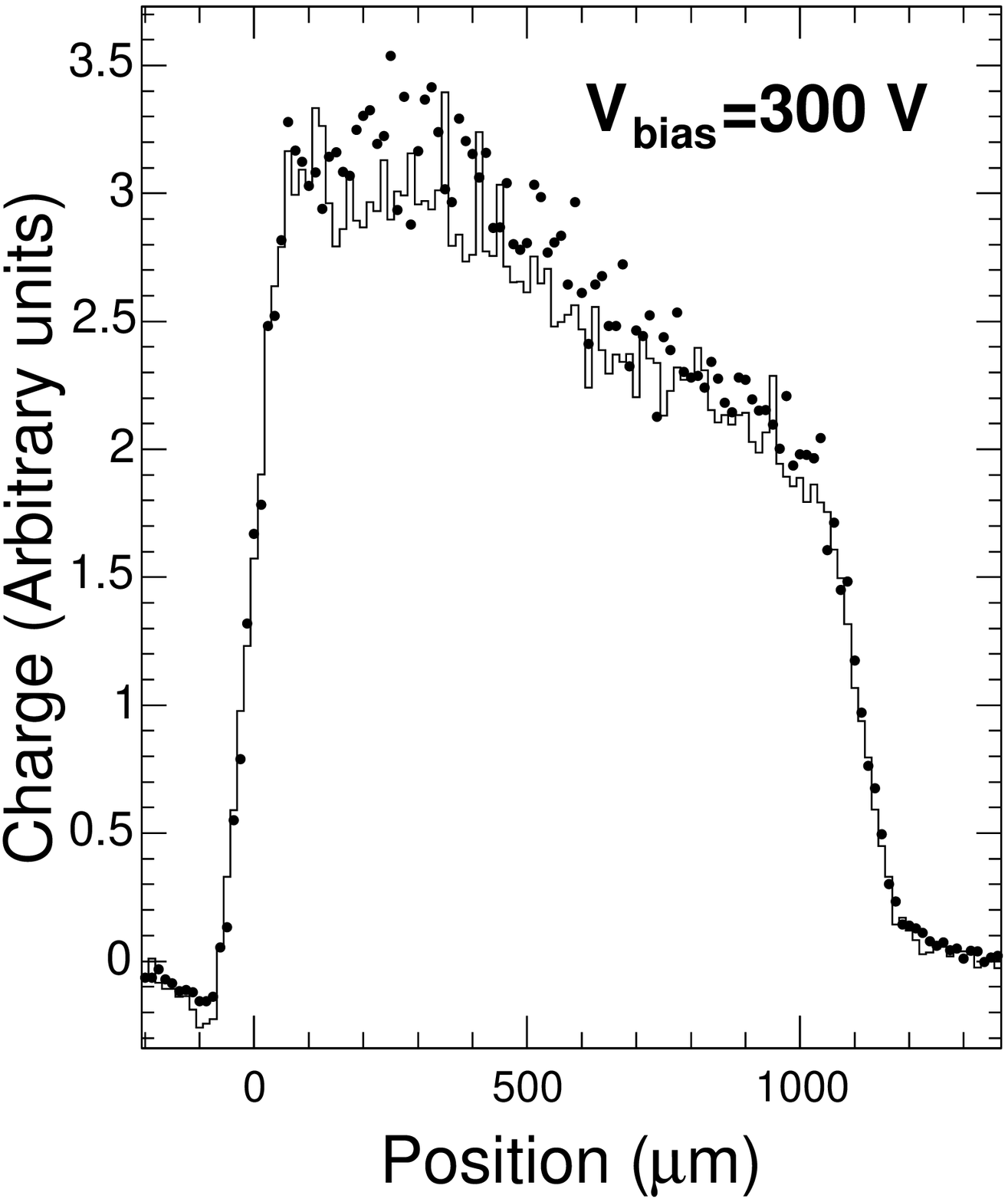,width=\textwidth}
	  \label{fig:dj44_c}
      }}
      \subfigure[]{\scalebox{0.24}{
	  \epsfig{file=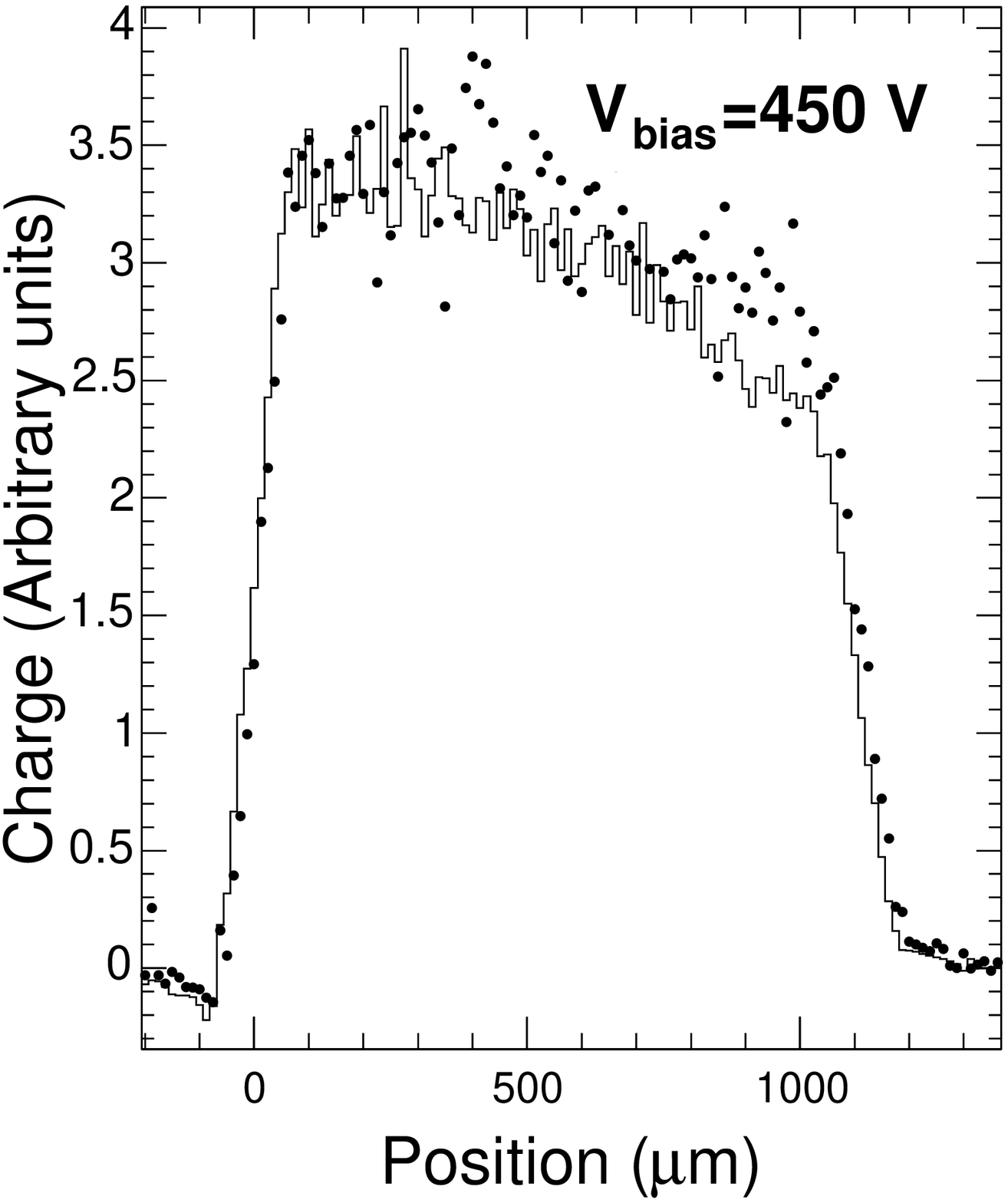,width=\textwidth}
	  \label{fig:dj44_d}
      }}
    }
  \caption{The measured charge collection profiles at bias voltages of 150 V, 200 V, 300 V, and 450 V are shown as solid dots 
    for fluences of $6\times10^{14}$~n$_{\rm eq}/$cm$^{2}$. 
    %and (b) $9.7\times10^{14}$~n$_{\rm eq}/$cm$^{2}$.  
    The BF simulation is shown as the solid histogram in each plot.}
  \label{fig:dj35} 
\end{figure*}

The measured charge collection profiles at bias voltages between 150 V and 450 V are compared with 
the BF simulation in Fig.~\ref{fig:dj35} for a fluence of $6\times10^{14}$~n$_{\rm eq}/$cm$^{2}$.
% and (b) $9.7\times10^{14}$~n$_{\rm eq}/$cm$^{2}$.  
The electron trapping rate for the lower fluence is set to 93\% of the nominal value.  
%The electron 
%trapping rate for the higher fluence sample which was held at 30$^\circ$C for a longer time is set 
%to 70\% of the nominal value.  
%The trap densities for the higher fluence simulation were scaled linearly 
%from the densities used for the lower fluence simulation.  

Although for high values of the bias voltage the simulation falls below the measured profile, it provides a reasonable description of the measurements.  Several features of the measured distributions are described well by the simulation.  Note that both data and simulations show a distinctly negative signal near $y=0\ \mu$m.  This can be understood as a consequence of hole trapping.  Electrons deposited near the n$^+$ implant are collected with high efficiency whereas holes deposited near the implant must transit the entire detector thickness to reach the p$^+$ backplane.  If the holes are collected, they produce no net signal on the n$^+$ side of the detector.  However if the holes are trapped, then a negative signal is induced and is most visible in the $y<0\ \mu$m region.  Another feature is the ``wiggle'' in the 150 V profiles.  The relative signal minimum near $y=700\ \mu$m corresponds to the $E_z$ minimum where both electrons and holes travel only short distances before trapping.  This small separation induces only a small signal on the n$^+$ side of the detector.  At larger values of $y$, $E_z$ increases causing the electrons drift back into the minimum where they are likely to trap.  However, the holes drift into the higher field region near the p$^+$ implant and are more likely to be collected.  The net induced signal on the n$^+$ side of the detector therefore increases and creates the local maximum seen near $y=900\ \mu$m.  

%At the $9.7\times10^{14}$~n$_{\rm eq}/$cm$^{2}$ fluence, the simulation provides a poorer description of the corresponding measurements than it does at the lower fluence.  It is likely that there was more annealing of the higher fluence sample and simple linear scaling of trap densities with fluence is probably not optimal.  The accommodation of the additional annealing may require an adjustment of the $N_A/N_D$ ratio or other degrees of freedom.  A second observation is that the measured and simulated signals are absolutely normalized and the signal observed at 450V from the high fluence sample is larger than that observed for the lower fluence sample.  This is surprising and may suggest that sample-to-sample normalization issues exist or it may confirm the idea that annealing has altered the behavior of the sample.

\begin{figure}[hbt]
  \begin{center}
    \resizebox{\linewidth}{!}{    
      \includegraphics[angle=90]{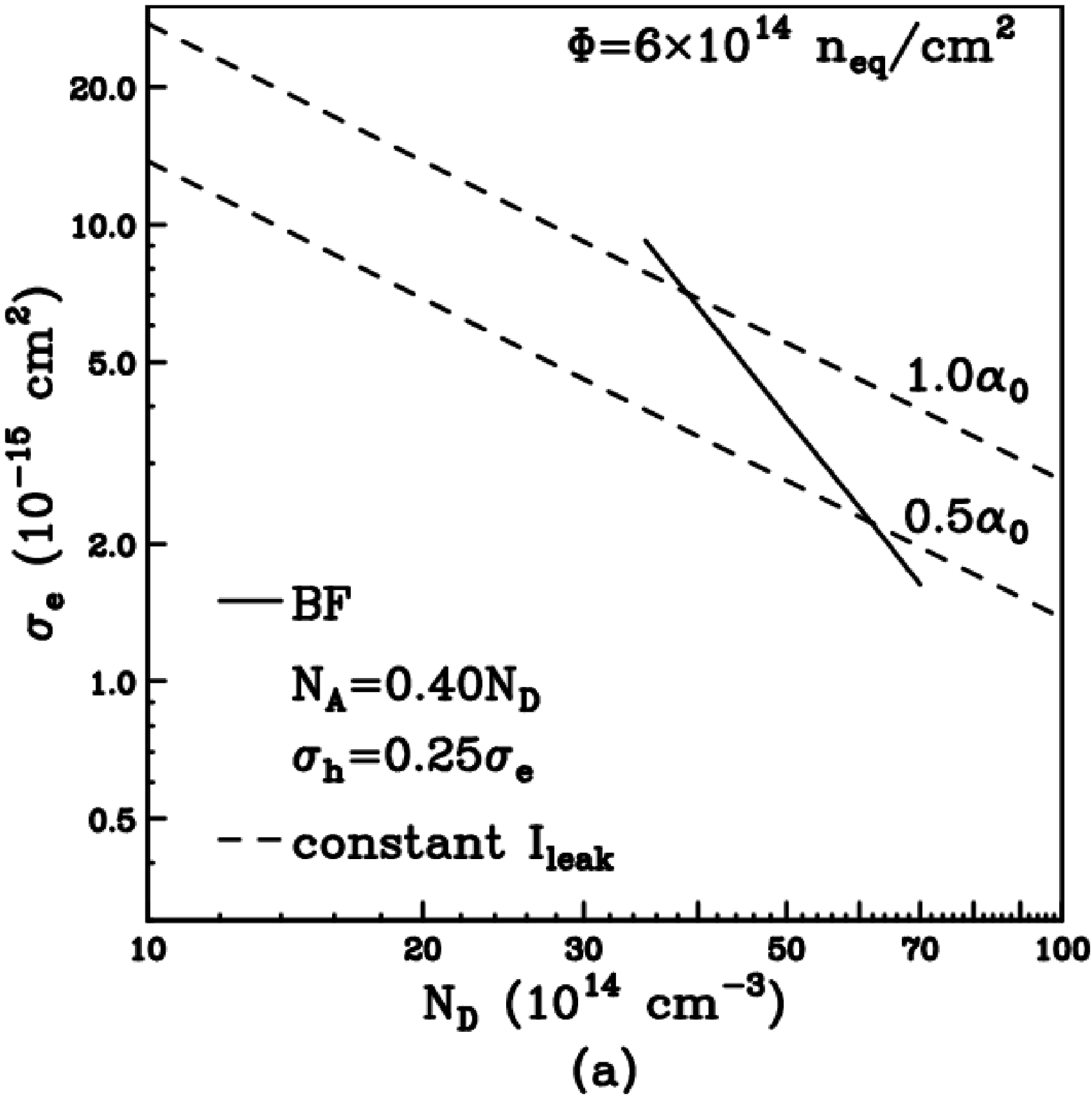} 
      \includegraphics[angle=90]{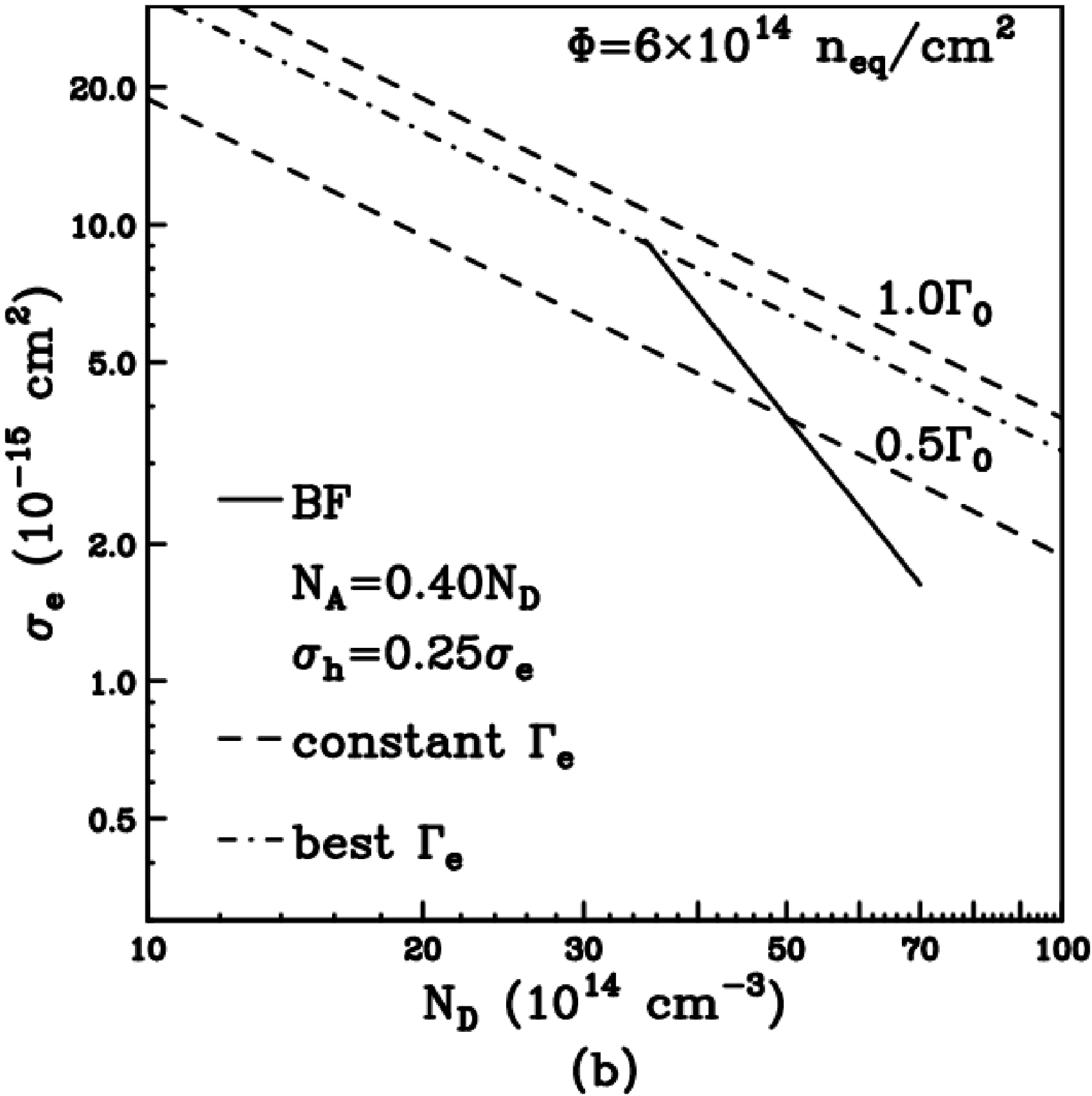} 
	      }
  \end{center}
  \caption{The allowed region in the $N_D$-$\sigma_e$ space for the BF model is shown as the solid line in (a) and (b).  Contours of constant leakage current are shown as dashed curves in (a) and are labeled in terms of the corresponding damage parameter $\alpha$ in units of $\alpha_0$.  Contours of constant electron trapping rate are shown as dashed curves in (b) and are labeled in terms of the un-annealed trapping rate $\Gamma_0$ for the nominal fluence.}
  \label{fig:dj35_parameters} 
\end{figure}

The BF model fixes the ratios $N_A/N_D$ and the ratio between cross sections $\sigma_h/\sigma_e$ leaving the parameters $N_D$ and $\sigma_e$ to vary.  Over a restricted range, an increase in $N_D$ can be offset by a decrease in $\sigma_e$ keeping the electric field profile approximately unchanged.  Scaling the electron cross section as $\sigma_e\propto N_D^{-2.5}$  produces very similar charge collection profiles.  The allowed region in the $N_D$-$\sigma_e$ space is shown in Fig.~\ref{fig:dj35_parameters}(a) as the solid line in the logarithmic space.  If the donor density becomes too small ($N_D<20\times10^{14}$~cm$^{-3}$), the 150 V simulation produces enhanced signal at large $z$.  If the donor density becomes too large ($N_D>50\times10^{14}$~cm$^{-3}$), the 300 V simulation produces insufficient signal at large $z$.  Since the simulated leakage current varies as $I_\mathrm{leak}\propto\sigma_e N_D$, different points on the allowed solid contour correspond to different leakage current.  Contours of constant expected leakage current are shown as dashed curves and are labeled in terms of the damage parameter $\alpha$ in units of $\alpha_0$.  It is clear that the simulation can accommodate the expected leakage current which is smaller than the measured current by a factor of three.

The electron and hole traps in the model should also contribute to the trapping of signal carriers.  The contributions of these states to the effective trapping rates of electrons and holes are given by the following expressions
\begin{eqnarray}
\Gamma_e  & = & v_e \left[\sigma_e^A N_A (1-f_A) + \sigma_e^D N_D f_D\right]  \simeq  v_e \sigma_e^A N_A \\
\Gamma_h  & = & v_h \left[\sigma_h^D N_D (1-f_D) + \sigma_h^A N_A f_A\right]  \simeq  v_h \sigma_h^D N_D 
\label{eq:signal_trap}
\end{eqnarray}
where it has been assumed that the trap occupancies are small and the thermal velocity of electrons at -10$^\circ$C is set to $v_e = 2.15\times10^7$ cm/s. Because $N_A/N_D$ is assumed to be constant, contours of constant electron trapping rate are parallel to contours of constant leakage current in $N_D$-$\sigma_e$ space.  The best ``fit'' of the simulation to the measured profiles reduced $\Gamma_e$ to 93\% of the un-annealed trapping rate $\Gamma_0 = \Phi \beta_e = 0.33$ ns$^{-1}$ for the nominal fluence $\Phi$~\cite{kramberger1}.  These contours are compared with the allowed contour in Fig.~\ref{fig:dj35_parameters}(b).  
It is clear that the simulation can accommodate the measured trapping rate in the same region of parameter space that maximizes the leakage current.

Figure~\ref{fig:dj35_parameters}(b) also suggests a solution to the puzzle that the trapping rates 
have been shown to be unaffected by the presence of oxygen in the detector bulk \cite{kramberger1} 
whereas it is well-established that the space charge effects are quite sensitive to the presence of oxygen in the material \cite{oxygen1,oxygen2}.  
It is clear from Fig~\ref{fig:dj35_parameters}(b) that small-cross-section trapping states can 
play a large role in the effective charge density but a small one in 
the effective trapping rates: every point on the BF line produces 
100\% of the effective charge density but only the larger cross section 
points contribute substantially to the trapping rate.  
%If we were to add additional trapping states with small cross sections and the 
%formation of those states were suppressed by oxygen, then $\rho_{eff}$ 
%would be sensitive to oxygenation whereas $\Gamma_{e/h}$ would be 
%insensitive to oxygenation.
%It is clear that additional small-cross-section trapping states can play a large role in the effective charge density but a small one in the effective trapping rates.  
If the formation of the additional small-cross-section states were suppressed by oxygen, then $\rho_\mathrm{eff}$ could be sensitive to oxygenation whereas $\Gamma_{e/h}$ would be insensitive to oxygenation.  
This is another consequence of the observation that the occupancies $f_{D/A}$ of the trapping states are independent of the scale of the cross sections in the steady state (see Section~\ref{sec:djmodels}).  The trapping of free carriers is not a steady-state phenomenon and is sensitive to the scale of the trapping cross sections.

%--------------------------------------------
% Conclusions
%--------------------------------------------
\section{Conclusions}
\label{sec:conclusions}

The main result of the work presented in this paper is that a double peak electric field is necessary to describe the charge collection profiles measured in heavily irradiated pixel sensors.  A simulation utilizing a linearly varying electric field based upon the standard picture of a constant type-inverted effective charge density is inconsistent with the measurements.

A two-trap EVL-like model can be tuned to provide a reasonable description of the measurements.  It can also account for the expected level of the leakage current (although not the observed leakage current) and the observed electron trapping rate.  It is important to state that any two-trap model is, at best, an ``effective theory''.  It is well-known that there are many trapping states in heavily irradiated silicon that trap charge.  There may also be thermodynamically ionized defects that contribute to the effective space charge density.  Clearly a two-trap model can describe the gross features of the physical processes in our sensors but it may not be able to describe all details.  This also implies that the parameters of the two-trap model presented in this paper are unlikely to have physical reality.  

The charge-sharing behavior and resolution functions of many detectors are sensitive to the details of the internal electric field.  A known response function is a key element of any reconstruction procedure.  A working effective model will permit the detailed response of these detectors to be tracked as they are irradiated in the next generation of accelerators.  

%Finally, we note that despite a growing body of contrary evidence, the overly simple picture of uniform type inversion in irradiated silicon detectors persists in the minds of many researchers and even in the terminology used by experts to describe their devices.  
%
Finally, we note that quantities like $W$ (depletion depth) and $N_\mathrm{eff}$, which are related to 
the picture of uniform type inversion in irradiated silicon sensors, may correctly suggest reduced 
detector performance but given the evidence of double peak electric fields and free carrier trapping, 
have no physical significance.
%
%After irradiation, quantities like $W$ (depletion depth) and $N_\mathrm{eff}$ may correctly suggest reduced detector performance but given the reality of double peak electric fields and free carrier trapping, they have no physical significance.  

%--------------------------------------------
% The Bibliography
%--------------------------------------------

\end{document}